\documentclass[aps,prd,superscriptaddress,twocolumn,showpacs,nofootinbib,preprintnumbers]{revtex4-1}

\usepackage{soul,color}
\usepackage{xspace,amsmath,amssymb,bm,bbm,dsfont,hyperref}
\usepackage{todonotes} 

\hypersetup{
 colorlinks=true,
 allcolors=[RGB]{31 119 180}
}
\usepackage{slashed}
\usepackage{multirow}
\DeclareMathOperator{\re}{Re}
\DeclareMathOperator{\im}{Im}

\newcommand{\diff}{\ensuremath{\mathrm{d}}}
\newcommand\bsub{\begin{subequations}}
\newcommand\esub{\end{subequations}}

\newcommand{\eg}{{\it e.g.}\xspace}
\newcommand{\ie}{{\it i.e.}\xspace}

\newcommand{\kev}{\ensuremath{{\mathrm{\,ke\kern -0.1em V}}}\xspace}
\newcommand{\mev}{\ensuremath{{\mathrm{\,Me\kern -0.1em V}}}\xspace}
\newcommand{\gev}{\ensuremath{{\mathrm{\,Ge\kern -0.1em V}}}\xspace}
\newcommand{\gevsq}{\ensuremath{{\mathrm{\,Ge\kern -0.1em V}^2}}\xspace}
\newcommand{\tev}{\ensuremath{{\mathrm{\,Te\kern -0.1em V}}}\xspace}

\newcommand{\M}[2]{\ensuremath{\mathcal{M}_{#1;#2}}\xspace}
\newcommand{\A}[2]{\ensuremath{\mathcal{A}_{#1;#2}}\xspace}

\newcommand{\betaVT}{\ensuremath{\beta^{\gamma T}_V}\xspace}
\newcommand{\betaVa}{\ensuremath{\beta^{\gamma a_2}_{V}}\xspace}
\newcommand{\betaVf}{\ensuremath{\beta^{\gamma f_2}_{V}}\xspace}
\newcommand{\betarhoa}{\ensuremath{\beta^{\gamma a_2}_{\rho}}\xspace}
\newcommand{\betarhof}{\ensuremath{\beta^{\gamma f_2}_{\rho}}\xspace}
\newcommand{\betaomegaa}{\ensuremath{\beta^{\gamma a_2}_{\omega}}\xspace}
\newcommand{\betaomegaf}{\ensuremath{\beta^{\gamma f_2}_{\omega}}\xspace}
\newcommand{\betaAT}{\ensuremath{\beta^{\gamma A}_T}\xspace}

\newcommand{\betaba}{\ensuremath{\beta^{\gamma a_2}_{b_1}}\xspace}
\newcommand{\betabf}{\ensuremath{\beta^{\gamma f_2}_{b_1}}\xspace}
\newcommand{\betaha}{\ensuremath{\beta^{\gamma a_2}_{h_1}}\xspace}
\newcommand{\betahf}{\ensuremath{\beta^{\gamma f_2}_{h_1}}\xspace}

\begin{document}

\title{Exclusive tensor meson photoproduction}

\newcommand{\cern}{CERN, 1211 Geneva 23, Switzerland}
\newcommand{\jlab}{Theory Center,
Thomas  Jefferson  National  Accelerator  Facility,
Newport  News,  VA  23606,  USA}
\newcommand{\ucm}{Departamento de F\'isica Te\'orica, Universidad Complutense de Madrid and IPARCOS, 28040 Madrid, Spain}
\newcommand{\icn}{Instituto de Ciencias Nucleares,
Universidad Nacional Aut\'onoma de M\'exico, Ciudad de M\'exico 04510, Mexico}
\newcommand{\ceem}{Center for  Exploration  of  Energy  and  Matter,
Indiana  University,
Bloomington,  IN  47403,  USA}
\newcommand{\indiana}{Physics  Department,
Indiana  University,
Bloomington,  IN  47405,  USA}
\newcommand{\ect}{European Centre for Theoretical Studies in Nuclear Physics and related Areas (ECT$^*$) and Fondazione Bruno Kessler, Villazzano (Trento), I-38123, Italy}
\newcommand{\genova}{INFN Sezione di Genova, Genova, I-16146, Italy}
\newcommand{\icsup}{Institute of Computer Science, Pedagogical University of Cracow, 30-084 Krak\'ow, Poland}

\author{V.~Mathieu}
\email{vmathieu@ucm.es}
\affiliation{\ucm}

\author{A.~\surname{Pilloni}}
\email{pillaus@jlab.org}
\affiliation{\ect}
\affiliation{\genova}

\author{M.~Albaladejo}
\affiliation{\jlab}

 \author{\L.~Bibrzycki}
 \affiliation{\jlab}
 \affiliation{\indiana}
 \affiliation{\icsup}
\author{A.~Celentano}
\affiliation{\genova}

\author{C.~\surname{Fern\'andez-Ram\'irez}}
\affiliation{\icn}

\author{A.~P.~Szczepaniak}
\affiliation{\jlab}
\affiliation{\indiana}
\affiliation{\ceem}

\collaboration{Joint Physics Analysis Center}
\begin{abstract}
We study tensor meson photoproduction outside of the resonance region, at beam energies of few Ge\!Vs. We build a model based on Regge theory that includes the leading vector and axial exchanges. We consider two determinations of the unknown helicity couplings, and 
fit to the recent $a_2$ photoproduction data from CLAS. Both choices give a similar description of the $a_2$ cross section, but result in different predictions for the parity asymmetries and the $f_2$ photoproduction cross section. 
We conclude that new measurements of $f_2$ photoproduction in the forward region are needed to pin down the correct production mechanism. We also extend our predictions to the $8.5\gev$ beam energy, where current experiments are running.
\end{abstract}

\preprint{JLAB-THY-20-3188}

\maketitle

\section{Introduction}
The lightest tensor meson multiplet is well established experimentally and theoretically~\cite{pdg,Jackura:2017amb,Dudek:2014qha,*Dudek:2016cru,*Briceno:2017qmb} and fits well into the quark model.  Given their relatively narrow width, light tensors 
 can be used as a benchmark when 
 searching for states which are less prominent in data, for example the $J^{PC}=1^{-+}$ exotic hybrid candidate~\cite{Adolph:2015tqa,Rodas:2018owy}.

A comprehensive understanding of tensor meson production dynamics  is thus needed to pin down the properties of hybrid 
mesons. 

 In particular, in  photoproduction both hybrids and tensors 
  can be produced through 
   vector and axial exchanges.   The $a_2(1320)^0$ 
   photoproduction cross section has been recently measured by the CLAS experiment in the $4$--$5\gev$ beam energy range~\cite{Celentano:2020ttj}. The $f_2(1270)$ cross section has not been extracted explicitly, but it can be inferred from the partial wave analysis of $\gamma p \to \pi^+\pi^- p$~\cite{Battaglieri:2009aa}.
A pattern seems to emerge from various photoproduction reactions: when isovector mesons 
like the $\pi^0$ or $a_2^0$ 
are produced, the differential cross section exhibits a dip at $t \simeq -0.5\gevsq$, which does not appear in photoproduction of isoscalars, 
like the $\eta$ or $f_2$. 

In this paper, we describe tensor meson photoproduction in the $3$--$10\gev$ beam energy range with a model based on Regge pole exchanges. The model is compared to CLAS data in Sec.~\ref{sec:vector}. In its simplest version, the amplitude includes the leading vector exchanges only, and leads to an exact zero at the so-called wrong-signature point.
The overall normalization is constrained from known tensor meson decay widths.
In Sec.~\ref{sec:axial}, we introduce axial exchanges as a possible mechanism to fill in the zero. The strength of vector and axial exchanges is refitted to the CLAS $a_2$ data. We then compare the predictions of the model with the $f_2$ cross section data.
Our predictions are extended to a 
higher beam energy
of $E_\gamma = 8.5\gev$, where GlueX and CLAS12 are currently operating~\cite{Ghoul:2015ifw,battaglieri2005meson}. Polarization observables sensitive to the naturality of the exchanges are predicted in Sec.~\ref{sec:obs}. Summary and conclusions are presented in Sec.~\ref{sec:conclusions}, while several technical details are left to the Appendices.

\section{Vector exchanges}\label{sec:vector}
\begin{figure}[b]
\begin{center}
	\includegraphics[width=0.6\linewidth]{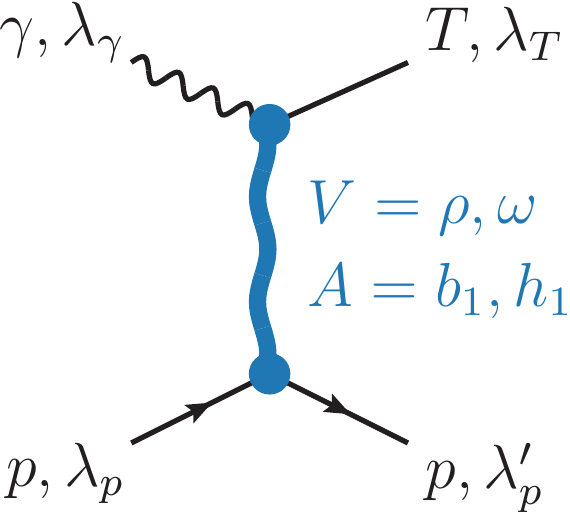}
\end{center}
\caption{\label{fig:reggeon}Factorization of the tensor meson
$T$
photoproduction amplitude {\it via} the Regge exchange 
$E=V,A$.}
\end{figure}
We consider the process $\gamma p \to T p$, where $T = a_2(1320)^0,\ f_2(1270)$.
We do not consider charge exchange processes, as $\gamma p \to a_2(1320)^+ n$ or $\gamma p \to K_2^*(1430)^+ \Lambda$  or $\to K_2^*(1430)^{+,0} \Sigma^{0,+}$, which are driven by pion or kaon exchange, respectively, and exhibit a different phenomenology  (for example, the former has been studied in~\cite{Wang:2015kia}). At high energies, the amplitude in the forward direction is
dominated by the leading Regge exchanges. As represented in Fig.~\ref{fig:reggeon}, Regge pole amplitudes factorize into a product of an upper and a lower vertex~\cite{Gribov:1962fw, Arbab:1969zr}, that describe the beam and target interactions, respectively. The amplitude can be written:

\begin{align}
	\M{\lambda_\gamma \lambda_T}{\lambda_p \lambda'_p} &= -\sum_E T^{E}_{\lambda_\gamma \lambda_T}(t) R^E(s,t)  B^E_{\lambda_p \lambda'_p}(t),
\end{align}
where $\lambda_i$ is the center-of-mass helicity of particle $i$, and  $s$, $t$ the Mandelstam variables describing the total energy squared, and the momentum transferred squared between the initial and final nucleon. The sum runs over the Regge poles that contribute to tensor production. As customary, 
reggeons are labelled by the lightest meson lying on the trajectory, and classified by 
this meson's quantum numbers, 
in particular parity $P$, signature $\tau = (-)^J$, and naturality $\eta = P(-)^J$. The dominant natural exchanges 
are the vector $\rho$ and $\omega$, while the unnatural ones are the axial $b_1$ and $h_1$. 
The beam asymmetry in $\pi^0$ and $\eta$ photoproduction by GlueX~\cite{AlGhoul:2017nbp} suggest that natural exchanges dominate over unnatural ones as long as pion exchange is forbidden~\cite{Mathieu:2015eia,Nys:2016vjz,Nys:2017xko}. For this reason, we now focus on the leading vector exchanges only, $V = \rho, \omega$. 

The Regge propagator is given by~\cite{Irving:1977ea}:
\begin{equation}\label{eq:ReggePropa}
	R(s,t) =  \frac{\tau +  e^{-i\pi\alpha(t)}}{2}  \, (-)^{\ell} \Gamma\left[\ell-\alpha(t)\right] (\alpha' s)^{\alpha(t)} .
\end{equation}
The factor $\Gamma\left[\ell-\alpha(t)\right]$ has poles for integers $\alpha(t)=J \ge \ell$, representing the exchange of a particle of spin $J$ in the $t$-channel. 
The signature factor $\tau +  e^{-i\pi \alpha(t)}$ cancels the wrong-signature poles at $J \ge \ell$, and provides additional wrong-signature zeroes for $J < \ell$. 

Duality arguments based on the nonexistence of flavor-exotic resonances, at least in the light sector, require the parameters in the propagator to be equal for vectors and tensors (exchange degeneracy, EXD). 
For the trajectories 
it holds 
since $\alpha(t) =   1 + \alpha'(t-m_V^2) \simeq 2 + \alpha'(t-m_T^2) \simeq \alpha' t+ 0.5$ with $\alpha' = 0.9\gev^{-2}$.  
The value of $\ell$ is the spin of the lightest state that appears on all the degenerate trajectories. Since there is no scalar meson on the leading trajectories, $\ell = 1$. For vector exchanges, $\tau = -1$ and the amplitude vanishes at $J=0$, which corresponds to $t = - 0.55\gevsq$. 
The propagator is normalized such that, at the vector pole:
\begin{align}
R(s, t \to m_V^2) & \to \frac{\alpha' s}{1- \alpha(t)}  = \frac{s}{m_V^2-t}.
\end{align}

The bottom vertex depends on two helicity couplings:
\begin{align}\nonumber
B^{V}_{\lambda_p \lambda'_p}(t) &= \left(\frac{-t'}{4m_p^2}\right)^{\frac{1}{2}|\lambda_p-\lambda'_p|} \\ &\qquad\times  \left[ G^{V}_1\delta_{\lambda_p,\lambda'_p}  + 2 \lambda_p G^V_2 \delta_{\lambda_p,-\lambda'_p} \right].
\label{eq:bottom}
\end{align}
The half-angle factor $(-t')^{\frac{1}{2}|\lambda_1-\lambda_2|}$ arises from conservation of angular momentum in the forward direction, with $t'$ defined as:
\bsub\begin{align}
t' &= t - t_\text{min} =  -4 q q' \sin^2 \frac{\theta}{2} = t - \frac{m_{T}^4}{4s} + (q-q')^2,
\end{align}\esub
with $q$, $q'$ the incoming and outgoing 3-momentum, $q = (s-m_p^2)\big/2\sqrt{s}$, $q'  = \lambda^{1/2}(s,m_p^2,m_T^2)\big/2\sqrt{s}$,
with $\lambda(a,b,c) = a^2+b^2+c^2-2(ab+bc+ca)$.

Similarly, in the top vertex, we factorize the half-angle factor and an overall normalization 
:
\begin{align} \label{eq:7}
 T^{V}_{\lambda_\gamma \lambda_T}(t) & = \betaVT \left(\frac{-t'}{m_T^2}\right)^{\frac{1}{2}|\lambda_\gamma-\lambda_T|}  \beta_{\lambda_\gamma \lambda_T}(t).
\end{align}
Parity conservation implies:
\begin{align}
     \beta_{-\lambda_\gamma \lambda_T}(t) = (-)^{\lambda_\gamma - \lambda_T}\beta_{\lambda_\gamma -\lambda_T}(t).\label{eq:parity}
\end{align}
The five independent helicity structures $\beta_{1,\lambda_T}(t)$ could, 
in principle, 
be extracted from the angular correlations of the decay $T \to V \gamma$. Unfortunately, these decay modes have not been measured yet. We thus must 
introduce a hypothesis to fix the relative size of the various structures, and fit the overall coupling to data. We consider two models, a ``minimal'' one (see \eg~\cite{Molina:2008jw,*Geng:2008gx,*Xie:2014twa,*Xie:2015isa}), and a second one that we refer to as Tensor Meson Dominance (TMD)~\cite{Suzuki:1993zs,*Giacosa:2005bw}. The helicity couplings $\beta_{\lambda_\gamma\lambda_T}$ of the two models are summarized in Table~\ref{tab:TVV}, and the derivation is described in Appendix~\ref{app:lagrangian}.
\begin{table}
\caption{Helicity structures $\beta_{\lambda_\gamma \lambda_T}(t)$ of the top vertex for the interaction models considered. The other structures can be obtained via the parity transformation in Eqs.~\eqref{eq:parity} and~\eqref{eq:parityunn}.\label{tab:TVV}}
\begin{ruledtabular}
\begin{tabular}{lccccc}
 & $\beta_{1,2}$ & $\beta_{1,1}$ & $\beta_{1,0}$ & $\beta_{1,-1}$ & $\beta_{1,-2}$  \\
\hline
Minimal & $0$ & $1/2$ & $-1/\sqrt{6}$ & $0$ &  $0 $ \\
TMD & $-1/2$ & $-t\big/2m_T^2$ & $t\big/2\sqrt{6} m_T^2$ &  $0$ & $0$\\ \hline
M1 & $0$ & $1/4$ & $-1/\sqrt{6}$ & $1/4$ & 0
\end{tabular}
\end{ruledtabular} \end{table}

\begin{figure*}
\begin{center}
	\includegraphics[width=\linewidth]{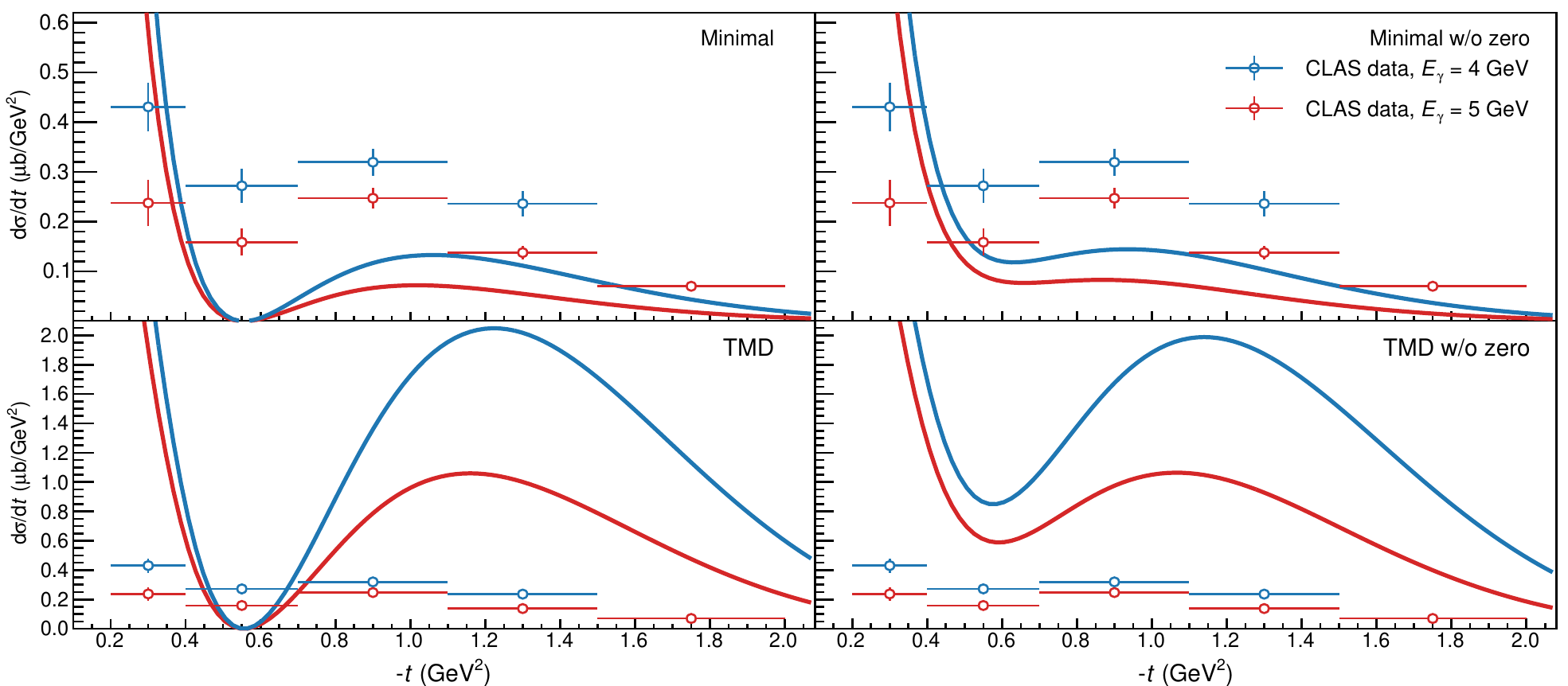}
\end{center}
\caption{\label{fig:onlyA2} Predictions for $a_2(1320)$ photoproduction differential cross section at $E_\gamma = 4$ (blue) and $5\gev$ (red). In the top panels we show the results for the minimal model, in the bottom ones the tensor meson dominance. 
The left plots feature the wrong-signature zero at $t = -0.55\gevsq$. In the right plots, we
modify the $\rho$ helicity-flip bottom coupling $G_2^\rho \to \frac{1}{\alpha(t)}G_2^\rho$ to fill in the zero,
as explained in the text.
The overall coupling is determined from the $\omega\pi\pi$ width. Data points from CLAS~\cite{Celentano:2020ttj}. 
}
\end{figure*} 

The overall normalization \betaVT could be extracted from the branching ratio of the radiative transitions between tensors and vectors. 
In the absence of this, we resort to Vector Meson Dominance (VMD), \ie we assume that the photon mixes with vector mesons through:
\begin{equation}
    {\cal L}_{VMD} =  - \sqrt{4\pi \alpha} A^\mu \left( 
m_\rho f_\rho\rho_\mu  +  m_\omega f_\omega \omega_\mu \right),
\end{equation}
where $f_{\rho,\omega}$ are the meson decay constants, and are related to the leptonic width $\Gamma\!\left(V \to e^+e^-\right) = 4\pi \alpha^2 f^2_V / m_V$. 

 Since the systematic uncertainties related to the model are much larger than the uncertainties of the parameters the model depends upon, we do not perform the usual error propagation, and just consider the qualitative behavior and the order of magnitude of these first estimates. 
 
In the following, we extract the couplings for the $a_2$, and leave the determination of the $f_2$ ones to Appendix~\ref{app:f2}.  We can use VMD to relate the radiative transition $T \to V\gamma$ to either the di-vector decay $T \to V V^{(\prime)}$, or the two-photon annihilation $T\to \gamma\gamma$. 
In the first method, we determine the \betaVa coupling from $\Gamma(a_2\to \omega\rho) \sim \Gamma(a_2\to \omega\pi\pi)  = 11.1\pm 3.4\mev$, assuming that the $\rho$ saturates the $\pi\pi$ pair~\cite{pdg}. 
The matrix element $\sum_\text{pol}|\mathcal{M}|^2$ is given in Eqs.~\eqref{width:TVV:Min} and~\eqref{width:TVV:TMD}, and must be averaged over the $\rho$ line shape:
\begin{align} \nonumber
    \Gamma(a_2\to \omega\rho) & = \frac{ (\beta_{a_2}^{\omega\rho})^2}{40\pi m_{a_2}^4} \int_{4 m_\pi^2}^{(m_{a_2}-m_\omega)^2} \frac{\diff s'}{\pi} \\
    &\quad\times \frac{\lambda^{1/2}\!\left(m_{a_2}^2, m_\omega^2, s'\right)}{2m_{a_2}} \sum_\text{pol}|\mathcal{M}|^2
    B_\rho(s'),
\end{align}
with
\bsub\label{eq:rhols}
\begin{align}
    B_\rho(s') & = \frac{ m_\rho \Gamma_\rho(s')}{(m_\rho^2-s')^2 + m_\rho^2 \Gamma_\rho^2(s')},
\\
    \Gamma_\rho(s') & = \Gamma_\rho \frac{m_\rho}{\sqrt{s'}} \left( \frac{s'-4m_\pi^2}{m_\rho^2-4 m_\pi^2} \right)^\frac{3}{2}.
\end{align}\esub
Finally, VMD leads to:
\begin{align}
    \betaVa & = \sqrt{4\pi \alpha} \frac{f_V}{m_V} \beta_{a_2}^{\omega\rho}.
\end{align}

Using the second method, we consider $\Gamma(a_2\to \gamma \gamma) = 1.00\pm 0.06\kev$~\cite{pdg}, and use Eqs.~\eqref{eq:2photon_min} and~\eqref{eq:2photon_TMD} to extract the two-photon couplings $\beta_{a_2}^{\gamma\gamma}$. With VMD, we obtain:

\begin{align}
    \betaomegaa & = \frac{\beta_{a_2}^{\gamma\gamma}}{\sqrt{4\pi \alpha}} \left( \frac{f_\omega}{m_\omega} + \frac{1}{3} \frac{f_\rho}{{m_\rho}} \right)^{-1}, &
    \betarhoa & =\frac{1}{3}\betaomegaa, \label{eq:fromgammatoomega}
\end{align}
using the isospin relations derived in Appendix~\ref{app:isospin}. The numerical values of the overall normalization obtained with these two methods for the two models studied are summarized in Table~\ref{tab:param}.

The differential cross section obtained using the di-vector decay width is compared to the CLAS data~\cite{Celentano:2020ttj} in Fig.~\ref{fig:onlyA2}. The model describes the dip in the $-t \in[0.4,0.6]\gevsq$ bin with an exact wrong-signature zero at $t=-0.55\gevsq$. To improve the agreement with data, we need to invoke a mechanism that partially fills in the zero.

 \begin{figure*}[t]
 \begin{center}
 	\includegraphics[width=\linewidth]{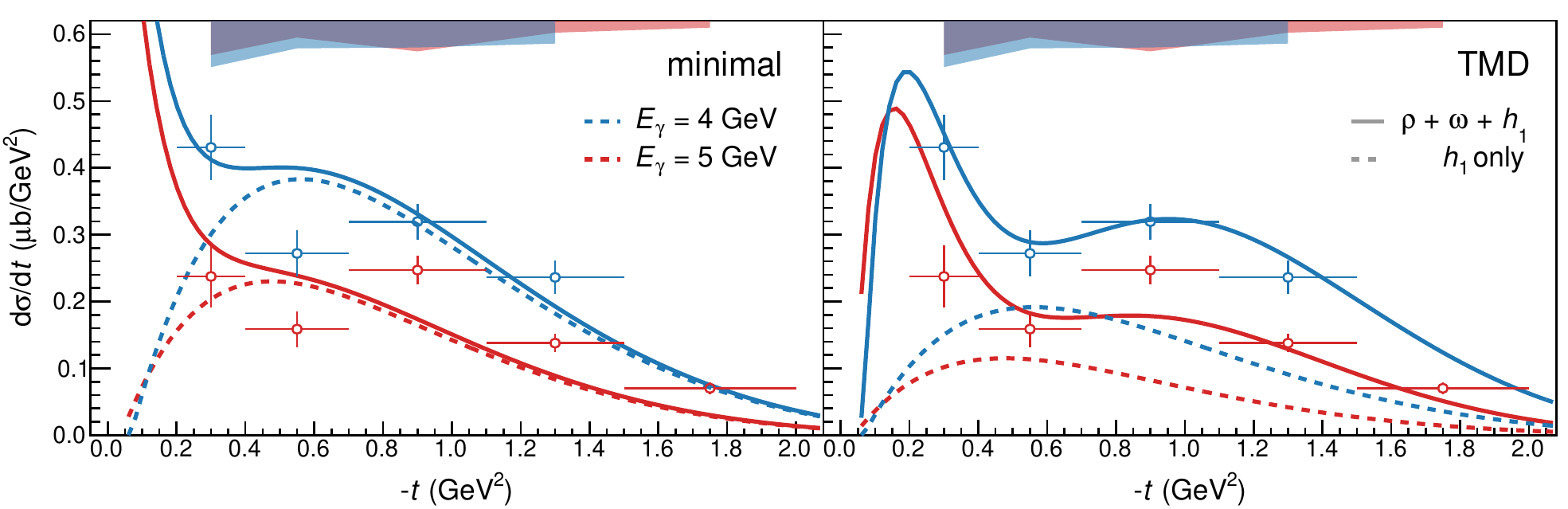}
 \end{center}
 \caption{\label{fig:fitted}Fit to $a_2$ photoproduction at $E_\gamma = 4$ and~$5\gev$. The minimal model (left panel) and TMD model (right panel) are fitted to the CLAS data~\cite{Celentano:2020ttj}. The solid lines show the full models, which includes both vector and axial exchanges. The strengths of vectors and axials is fitted to data. The contribution of axials is shown separately with dashed lines. The systematic uncertainties of~\cite{Celentano:2020ttj} are reported in the bands on top, and have not been considered in the fit.
}
 \end{figure*}

There is phenomenological evidence that the $\rho$ nucleon helicity-flip amplitude does not have the wrong-signature zero~\cite{Mathieu:2015gxa,Mathieu:2018mjw}. 
For example in $\eta$ photoproduction, which is dominated by $\rho$ exchange, the cross section does not dip~\cite{Nys:2016vjz}.
Accordingly, we will modify the helicity-flip bottom coupling $G_2^\rho \to \frac{1}{\alpha(t)}G_2^\rho$ to remove the wrong-signature zero. The predicted curves are shown in the right panels of Fig.~\ref{fig:onlyA2}. 

Both models, in particular the minimal one, have roughly the right order of magnitude. However, they fail at giving a good description of data. Moreover, from Table~\ref{tab:param} we notice that the \betaVT obtained using VMD from different reactions are substantially different. In the next section we will refit the overall normalization to data.

\section{Unnatural exchanges and comparison with $f_2(1270)$ data}\label{sec:axial}
\begin{figure*}[t]
\begin{center}
	\includegraphics[width=\linewidth]{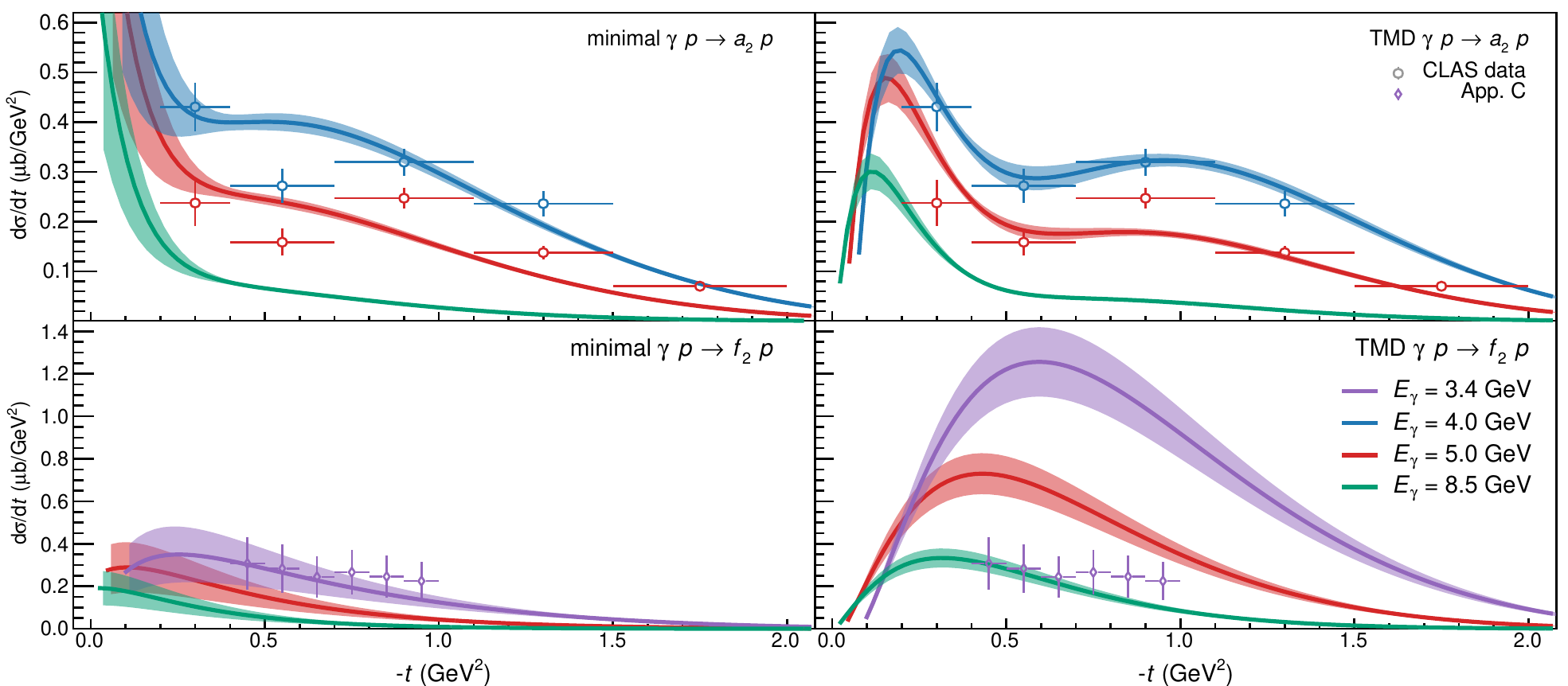}
\end{center}
\caption{\label{fig:prediction} Differential cross sections of $a_2$ (top) and $f_2$ (bottom panels), for different beam energies. The minimal model is shown in the left panel, the TMD in the right ones. The strengths of vectors and axials are fitted to the $a_2$ data only. The results are shown in Table~\ref{tab:fit}.
The error bands show the $1\sigma$ confidence interval which results from the statistical uncertainty of the fit. The $a_2$ data are taken from CLAS~\cite{Celentano:2020ttj}, the extraction of the $f_2$ data from the partial wave analysis of $\pi^+\pi^-$ by CLAS~\cite{Battaglieri:2009aa} is described in Appendix~\ref{app:f2xsect}. 
}
\end{figure*}

One can wonder whether other exchanges contribute to filling in the zero. 
If the strength of the dip is due to the nonflip $\rho$ exchange only, the isospin relations (given in Appendix~\ref{app:isospin}) predict that the $f_2$ cross section is nine times larger than the $a_2$ one at the wrong-signature point. On the contrary, they are comparable, as one can see from Fig.~\ref{fig:prediction}.
This suggests 
the existence of other isoscalar exchanges that contribute to filling in the zero. Isoscalar axial exchanges play a significant role in $\pi^0$ and $\eta$ photoproduction~\cite{Nys:2016vjz, Mathieu:2018mjw}. We investigate here how much they are relevant in tensor photoproduction.


The Regge propagator for axials is given by Eq.~\eqref{eq:ReggePropa} with $\ell = 0$, since the lowest spin on the EXD trajectory is the pion. The unnatural Regge trajectory is $\alpha(t) = \alpha'(t-m_\pi^2)$, with $\alpha'=0.7\gev^{-2}$.
Charge conjugation invariance restricts
the bottom vertex to the helicity-flip component only,
\begin{align}
    B^A_{\lambda_p \lambda'_p}(t) & = G^A_2 \left( \frac{-t}{4 m_p^2}\right)^{\frac{1}{2}|\lambda_p-\lambda_p'|}  \delta_{\lambda_p,-\lambda'_p},
\end{align}
with the coupling obtained from Ref.~\cite{Irving:1977ea}, $G_2^A = 25.24$, taking into account the normalization properly. The top vertex reads
\begin{align}
    T^A_{\lambda_\gamma \lambda_T}(t) & = \betaAT \left( \frac{-t}{ m_T^2}\right)^{\frac{1}{2}|\lambda_\gamma-\lambda_T|}  \beta_{\lambda_\gamma,\lambda_T},
\end{align}
with parity conservation implying 
\begin{align}
    \beta_{-\lambda_\gamma,\lambda_T} & = -(-)^{\lambda_\gamma - \lambda_T} \beta_{\lambda_\gamma,-\lambda_T}.\label{eq:parityunn}
\end{align}
In the absence of information on the angular distributions of the $T \to A \gamma$ decay, we restrict ourselves to the M1 transition that dominates in the nonrelativistic quark model. This fixes the relative size of the various helicity structures (see details in Appendix~\ref{app:lagrangian}), reported in Table~\ref{tab:TVV}.

Transitions of tensors to axials have not been observed, so we cannot proceed in the same way as we did for the natural case to predict the couplings. Moreover, from Table~\ref{tab:param} we notice that the \betaVT obtained using VMD from different reactions are substantially different.  Therefore, we now refit both vector and axial couplings to the $a_2$ data. 
We notice that the amplitude of $h_1$ and $b_1$ are identical, thus the fit is sensitive to the sum of couplings $\betaba + \betaha$ only. We know that $h_1$ and $b_1$ contribute equally to $\eta$ photoproduction, and that the former is nine times larger than the latter in $\pi^0$ photoproduction~\cite{Nys:2016vjz,Mathieu:2018mjw}. This agrees with the expectation from the isospin relations discussed in Appendix~\ref{app:isospin}. We thus set:
\begin{align}\label{eq:BetaFit}
    \beta^V & = \betaomegaa = 3\betarhoa, & 
    \beta^A & = \betaha, 
    & \betaba = 0.
\end{align}

\begin{table}
\caption{Parameters extracted from known decay widths. The bottom vertex couplings are taken from~\cite{Irving:1974ak}. \label{tab:param}}
\begin{ruledtabular}
\begin{tabular}{llcccc}
 & & \betarhoa & \betaomegaa & \betarhof & \betaomegaf  \\
\hline
\multirow{2}{*}{$\Gamma_{VV'}$}
& Minimal & $0.235 $ & $0.791$ & $0.700$ & $0.233$ \\
& TMD & $1.143$ & $3.8373$ & $3.31822$ & $1.10607$ \\ \multirow{2}{*}{$\Gamma_{\gamma\gamma}$}
& Minimal  & $0.110 $ & $0.331$ & $0.316$ & $0.105$ \\
& TMD & $0.238 $ & $0.715$ & $0.684$ & $0.228$ \\ 
\hline\hline
& & $G_1^\rho$ & $G_2^\rho$ & $G_1^\omega$ & $G_2^\omega$ \\ \hline
\multicolumn{2}{l}{Bottom vertex} & $1.63$ & $13.01$ & $8.13$ & $1.86$00
\end{tabular}
\end{ruledtabular}
\end{table}

\begin{table}
\caption{Fitted 
couplings 
defined in Eq.~\eqref{eq:BetaFit}. The error quoted is statistical and results from the fit.\label{tab:fit}}
\begin{ruledtabular}
\begin{tabular}{lcc}
 & $\beta^V$ & $\beta^A$  \\
\hline
Minimal & $0.251 \pm 0.053$ & $0.821\pm 0.023$ \\
TMD & $1.060\pm0.073$ & $0.581\pm0.053$
\end{tabular}
\end{ruledtabular} \end{table}

\begin{figure*}[t]
\begin{center}
	\includegraphics[width=\textwidth]{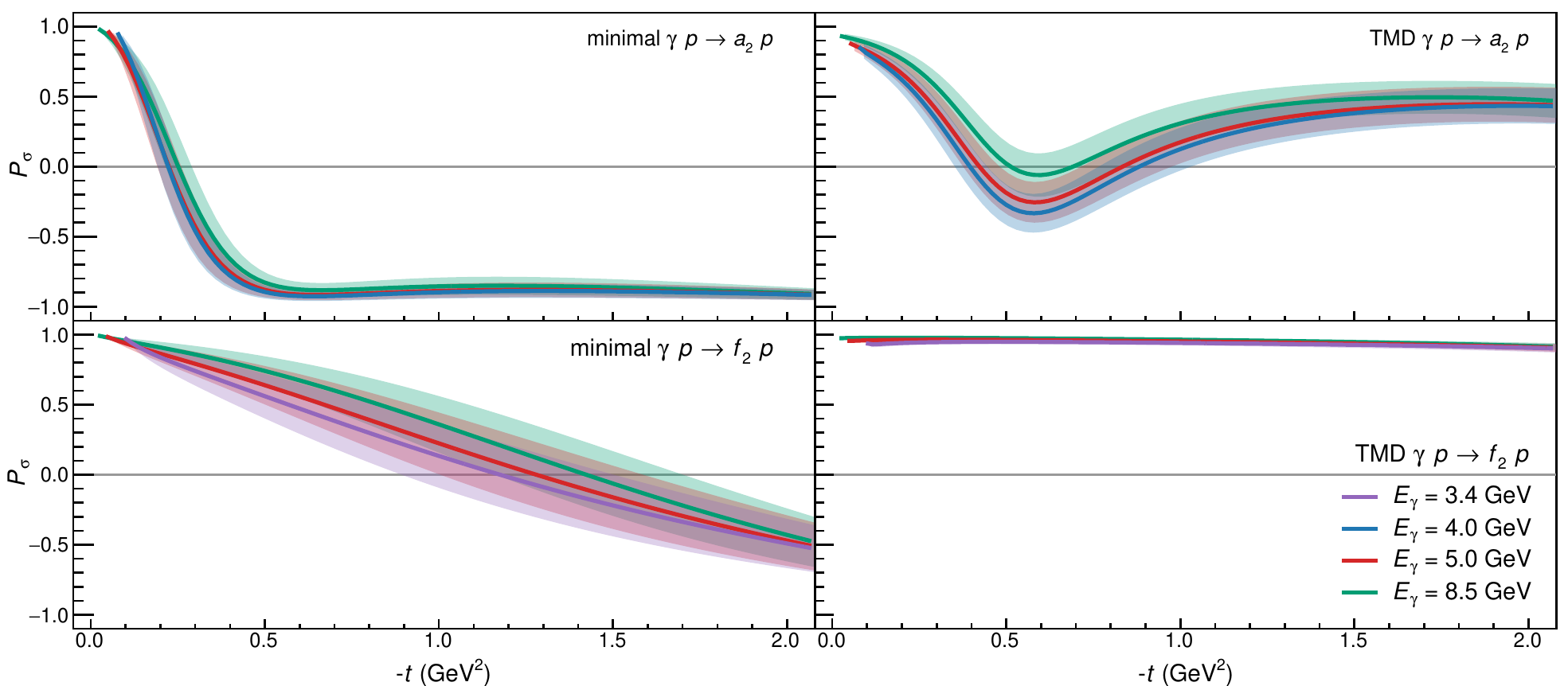}
\end{center}
\caption{\label{fig:Psigma} Parity asymmetry $P_\sigma$ in $a_2$ and $f_2$ photoproduction, for different beam energies. The minimal model is shown in the left panel, the TMD in the right ones. The strengths of vectors and axials are fitted to the $a_2$ differential cross section data only. The results are shown in Table~\ref{tab:fit}.
The error bands show the $1\sigma$ confidence interval which results from the statistical uncertainty of the fit.
}
\end{figure*}
We fit these two overall normalizations to data, using the sets of helicity structures given in Table~\ref{tab:TVV}. The systematic uncertainties from~\cite{Celentano:2020ttj} are not considered in the fit. The results are shown in Fig.~\ref{fig:fitted} and the fit parameters are summarized in Table~\ref{tab:fit}. We quote the statistical uncertainty on the fit parameters which propagates from the statistical uncertainties of the data points. 
No significant difference between the two models appears for $-t \gtrsim 0.6\gevsq$. In the forward region, the TMD model vanishes quickly due to the presence of higher derivatives, which turns into an additional factor of $t$ in the nonflip $\beta_{1,1}$ helicity coupling. It is also worth noting that, away from the very forward region, the cross section is dominated by unnatural exchanges. This is not the case for the TMD model, which captures  the wrong-signature dip at $t = -0.55\gevsq$ much better. More data at $-t \lesssim 0.6\gevsq$ will help in discriminating between the two models.

We discuss now $f_2$ photoproduction. Using the isospin relations of Appendix~\ref{app:isospin}, 
we can predict the behavior of the differential cross section. 
If we include the $h_1$ exchange only, the ratio of the couplings is $\betaha/\betahf = 3$. 
If instead we set $\betabf = \betahf$ as suggested by $\eta$ photoproduction~\cite{Nys:2016vjz}, the ratio of the contribution from axials to $a_2$ and to $f_2$ amplitudes would be $(\betaba + \betaha)/(\betabf + \betahf) = 5$. Since we are fitting the $a_2$ data only, this choice affects the predictions of the $f_2$. Moreover, since the $f_2$ cross section is dominated by $\rho$ exchange, choosing either value of $\betabf$ makes little difference. In the following, we show the results for $\betaba = \betabf = 0$.

Information about the $f_2$ may be extracted from the CLAS partial wave analysis of $\pi^+\pi^-$ photoproduction~\cite{Battaglieri:2009aa}. Since data are available in bins of $t$ and beam energy, we get the $f_2$ from the $\pi\pi$ $D$-wave with a simple fit as described in Appendix~\ref{app:f2xsect}. A new analysis by CLAS, dedicated to the $f_2$ cross section extracted from $\pi^0\pi^0$ photoproduction, is currently ongoing and will be published soon~\cite{CLASprivate}. We notice two main features: 
data look much flatter in $t$, and 
there is no evidence of the wrong-signature dip. 
As seen in Fig.~\ref{fig:prediction}, the minimal and TMD models differ significantly. We already noticed that the TMD vanishes in the forward direction, in opposition to the minimal one. Moreover, the former peaks at $t\simeq-  0.6\gevsq$, while the latter at $t\simeq - 0.2\gevsq$. These differences persist at higher beam energies. The minimal model fits well the $f_2$ data, while the TMD overshoots the data by a factor of 4.
In Fig.~\ref{fig:prediction} we also show the predictions for cross sections and parity asymmetries at the beam energy $E_\gamma = 8.5\gev$, which will be measured soon by GlueX and CLAS12.

Finally, we would like to comment about $f_2'(1525)$ production, which could be extracted from a $\gamma p \to K^+ K^- p$ partial wave analysis. In the ideal mixing scenario, the leading exchanges are  $\phi$ and $h_1'(1415)$. The formalism is identical to the one discussed above. Since the couplings are independent from the $a_2$ and $f_2$ ones, and there are no data to fit, we cannot provide reliable predictions.

\section{Polarization observables}\label{sec:obs}
The GlueX experiment operates with a linearly polarized beam at peaking energy $E_\gamma = 8.5\gev$. The photon polarization can also be extracted at the CLAS12 experiment, by measuring the angular distribution of the impinging electron. 
This information, correlated with the angular distribution of the tensor meson decay products, allows to extract the spin density matrix elements (SDME).
From the latter, we construct the parity asymmetry $P_\sigma$, which measures the relative strength of vector and axial exchanges: the asymmetry is close to $1$ when the natural exchanges dominate, and to $-1$ when the unnatural exchanges dominate. The definitions of SDME and $P_\sigma$ are given in Appendix~\ref{app:obs}. We present in Fig.~\ref{fig:Psigma} the predicted behavior of $P_\sigma$ for $a_2$ and $f_2$ photoproduction.

The predictions of $a_2$ parity asymmetry in the minimal and TMD models differ substantially. The dominance of axial exchanges for $-t\gtrsim 0.4\gevsq$ drives $P_\sigma$ towards $-1$ in the minimal model, while in the TMD model the parity asymmetry stays positive. For the $f_2$, the importance of axial exchanges grows as $-t$ increases in the minimal model, while the dominance of $\rho$ exchange in the TMD model for the $f_2$ leads to a parity asymmetry close to $1$.

\section{Conclusions}
\label{sec:conclusions}
In this paper we studied tensor meson photoproduction in the $3$--$10\gev$ beam energy range, based on a
Regge model, with vector and axial exchanges. We considered   
two different schemes for the vector helicity couplings. We first give an order-of-magnitude estimate of the couplings in both models. We then fit the $a_2$ data recently published by CLAS~\cite{Celentano:2020ttj}. We  predicted 
the $f_2$ cross section and compared to $f_2$ data extracted from a partial wave analysis of $\pi^+\pi^-$ photoproduction, also by CLAS~\cite{Battaglieri:2009aa}. While the two models give similar descriptions 
of the $a_2$ cross section, they differ in predicting the parity asymmetries and the $f_2$ cross section. The so-called minimal model provides better overall agreement with both $a_2$ and $f_2$ data, but at the price of missing the dip in $a_2$. Moreover, it predicts that the $a_2$ cross section is dominated by unnatural exchanges, which is at odds with the phenomenology of single meson photoproduction. On the other hand, the TMD model appears better grounded phenomenologically, but it overestimates the $f_2$ data. New data on both $a_2$ and $f_2$ photoproduction cross sections and beam asymmetries, in particular in the $-t \lesssim 0.6\gevsq$ region
, will allow us to pin down the exact strength of vector and axial contributions, and lead to a better understanding of the tensor meson production mechanisms.
The code to reproduce these results can be 
accessed at the JPAC website~\cite{JPACweb}.

\acknowledgments
We thank Ken Hicks for useful discussions.
This work was supported by the U.S.~Department of Energy under Grants
No.~DE-AC05-06OR23177 
and No.~DE-FG02-87ER40365, 
by  PAPIIT-DGAPA (UNAM, Mexico) under Grant No.~IA101819, 
and by CONACYT (Mexico) under Grant 
No.~A1-S-21389, 
by Polish Science Center (NCN) Grant No.2018/29/B/ST2/02576, 
and by the Istituto Nazionale di Fisica Nucleare (Italy). 
V.M. is supported by the Comunidad Aut\'onoma de Madrid through the Programa de Atracci\'on de Talento Investigador 2018 (Modalidad 1). 
\appendix

\section{Top vertex models}
\label{app:lagrangian}

\subsection{Photon-tensor-vector interaction}
The parity-conserving interaction between a tensor and two vectors involves 5 independent Lorentz structures. In the decay kinematics $T(t^{\mu\nu}, p_1+p_2) \to V_1 (\epsilon^{(1)}, p_1) + V_2 (\epsilon^{(2)}, p_2) $, the most generic covariant amplitude takes the form~\cite{Levy:1975fk}:
\begin{align} \nonumber
\mathcal{M}&=\frac{\beta^{V_1 V_2}_{T}}{m_T} t^{\mu\nu} \bigg[ 
  \alpha \epsilon^{*(1)}_\mu  \epsilon^{*(2)}_\nu 
 + \beta_1 (\epsilon^{*(1)}\cdot p_2)  \epsilon^{*(2)}_\mu p_{1\nu} \\ \nonumber
 &\quad
 + \beta_2 (\epsilon^{*(2)}\cdot p_1)  \epsilon^{*(1)}_\mu p_{2\nu} 
  + \gamma (\epsilon^{*(1)}\cdot \epsilon^{*(2)}) p_{1\mu} p_{2\nu} \\
  & \quad  + \delta (\epsilon^{*(1)}\cdot p_2) (\epsilon^{*(2)}\cdot p_1) p_{1\mu} p_{2\nu} 
 \bigg],
 \label{eq:TVV}
\end{align}
which leads to the decay width:
\begin{align}\label{eq:width}
 \Gamma(T\to V_1 V_2)  = \frac{\left(\beta^{V_1 V_2}_{T}\right)^2}{40 \pi } \frac{p}{m_T^4}\sum_{\lambda_1\lambda_2} |\mathcal{M}_{\lambda_1\lambda_2}|^2,
\end{align}
with $p = \lambda^{1/2}(m_T^2,m_1^2,m_2^2)/2m_T$, $E_i = \sqrt{p^2 + m_i^2}$, and:
\bsub\begin{align} 
\mathcal{M}_{11} &= \frac{\alpha - 2p^2 \gamma}{\sqrt{6}},
\\
\mathcal{M}_{10} &= \frac{p^2 m_T \beta_2 + E_2 \alpha }{m_2 \sqrt{2}},
\\
\mathcal{M}_{01} &= \frac{p^2 m_T \beta_1 + E_1 \alpha }{m_1 \sqrt{2}},
\\
\mathcal{M}_{00} &= \frac{\sqrt{2/3}}{m_1 m_2} 
  \big[ E_1 E_2 \alpha + p^2 m_T (E_2 \beta_1 + E_1 \beta_2)\nonumber\\ &\quad+ 
    p^2(E_1E_2 + p^2) \gamma + p^4 m_T^2 \delta\big],
    \\
\mathcal{M}_{1-1} &= \alpha, 
\\
\mathcal{M}_{-\lambda_1 -\lambda_2} &= \mathcal{M}_{\lambda_1 \lambda_2}.
\end{align}
\esub

In order to extract the Regge couplings from Eq.~\eqref{eq:TVV}, we write the amplitude of the process $\gamma(\lambda_\gamma) \gamma(\lambda_\gamma') \to T(\lambda_T)T(\lambda_T')$ with vector exchange in the $t$ channel, at leading order in $s$. By matching to the expected form 
[Eq.~\eqref{eq:7}]:
\begin{align} \nonumber 
    A_{\lambda_\gamma\lambda_T, \lambda_\gamma'\lambda_T'} & = \left(\beta^{\gamma T}_{V}\right)^2 \beta^V_{\lambda_\gamma,\lambda_T} 
    \left(\frac{-t}{m_T^2}\right)^{\frac{1}{2}|\lambda_\gamma-\lambda_T|} \\
    & \times \frac{s}{m_V^2-t}  \left(\frac{-t}{m_T^2}\right)^{\frac{1}{2}|\lambda_\gamma'-\lambda_T'|}
    \beta^V_{-\lambda_\gamma',-\lambda'_T},
\end{align}
we obtain the structures for the Regge couplings:
\bsub\begin{align}
    \beta_{1,2} & =   (2\beta_2 - t \delta)/4, 
    \\
    \beta_{1,1} & =  \frac{1}{4m_T^2} \big[  (2t+m_T^2)\beta_2-t\beta_1 \\&\quad -t(t+m_T^2) \delta  + 2\alpha \big],
    \\
    \beta_{1,0} & =  \frac{-1}{ 4\sqrt{6} m_T^2} \Big[4\alpha - 2(m_T^2+t)\beta_1 +2(t+2m_T^2) \beta_2\nonumber \\
    &\quad  -(t^2+4 t m_T^2+m_T^4) \delta \Big], 
    \\
    \beta_{1,-1} & =   \left[\beta_2 -\beta_1 - (t+m_T^2) \delta\right]/4,
    \\
     \beta_{1,-2}& = m_T^2 \delta/4,
     \\
      \beta_{-1,\lambda_T} &= (-)^{1 - \lambda_T}\beta_{1, -\lambda_T}.
\end{align} \esub
So far, the equations are completely generic, since no assumption has been made on the $(\alpha,\beta_1,\beta_2,\gamma,\delta)$ scalar functions. We also remark that the amplitude in Eq.~\eqref{eq:TVV} is not automatically gauge invariant when $V_1$ is massless. 
In the absence of information about the multipoles of $T \to V \gamma$, we consider two possible models, described below.

\subsection{The minimal model}
The ``minimal'' model is inspired by effective field theories (EFTs), and prescribes
to neglect all the terms with particle momenta, which correspond to higher derivative interactions in the EFT 
Lagrangian~\cite{Xie:2015isa}. We thus set $\alpha = m_T^2$ and $\beta_1 = \beta_2 = \gamma = \delta = 0$. The resulting covariant amplitude is not explicitly gauge invariant, so one needs to restrict manually the sum in Eq.~\eqref{eq:width} to the transverse photon polarizations. The widths read:
\bsub\begin{align} \label{width:TVV:Min}
 \Gamma(T\to V_1 V_2) &= \frac{p}{8\pi}\left(\beta^{V_1V_2}_{T}\right)^2 \Bigg[ 1 + \frac{1}{3}p^2 \left(\frac{1}{m_2^2} + \frac{1}{m_1^2}\right)\nonumber \\ &\qquad+ \frac{2}{15} \frac{p^4}{m_1^2 m_2^2}\Bigg],
 \\
  \Gamma(T\to \gamma \gamma)  &= \frac{7 m_T\left(\beta^{\gamma\gamma}_{T}\right)^2}{480\pi},
  \label{eq:2photon_min}
\end{align}\esub
while the Regge structures are reported in Table~\ref{tab:TVV}.

\subsection{Tensor Meson Dominance}
Tensor Meson Dominance (TMD)~\cite{Suzuki:1993zs} assumes that a tensor meson couples 
to a vector field with the stress-energy tensor, $\mathcal L = T^{\mu\nu} F_{\mu \rho} F_\nu^\rho$. The coupling to two distinct vectors is easily achieved by considering two distinct curvature tensors. The 
Lagrangian is manifestly gauge invariant. This model corresponds to setting $\alpha = p_1 \cdot p_2$, $\gamma = -\beta_1 = -\beta_2 = 1$, and $\delta = 0$ in Eq.~\eqref{eq:TVV}. The widths read:
\bsub\begin{align} \label{width:TVV:TMD}
 \Gamma(T\to V_1 V_2) &= \frac{p}{8\pi m_T^4}\left(\beta^{V_1V_2}_{T}\right)^2 \Bigg[m_1^2 m_2^2 \nonumber \\ &\quad +\frac{p^2}{3}  (m_T^2 + m_1^2 + m_2^2) + \frac{4}{15}p^4 \Bigg],
 \\
 \Gamma(T\to \gamma\gamma) &= \frac{m_T}{320\pi}\left(\beta^{\gamma\gamma}_{T}\right)^2,
 \label{eq:2photon_TMD}
\end{align}\esub 
while the Regge structures are reported in Table~\ref{tab:TVV}.

\subsection{Photon-tensor-axial interaction}
The parity-conserving interaction between a tensor, an axial and a vector involves 4 independent Lorentz structures:
\begin{align}
\mathcal{M} &= -i \frac{\beta^{VT}_A}{m_T} \varepsilon^{\mu\nu\rho\sigma} t^\alpha_\mu \left(p_V + p_A\right)_\sigma \Big[ 
\alpha_1 \, \epsilon^{*V}_\alpha \epsilon^{*A}_\nu p_{V\rho} \nonumber\\
&\quad+\beta_1 \, p_{V\alpha} \epsilon^{*A}_\nu p_{V\rho} \, \epsilon^{*V}\cdot p_A +\alpha_2 \, \epsilon^{*A}_\alpha \epsilon^{*V}_\nu p_{V\rho} \nonumber\\
&\quad+\beta_2 \, p_{V\alpha} \epsilon^{*V}_\nu p_{V\rho} \, \epsilon^{*A}\cdot p_V 
\Big].
\end{align}
However, since in the nonrelativistic quark model the transition $T(t^{\mu\nu}, p_\gamma +p_A) \to \gamma (\epsilon^{\gamma}, p_\gamma) + A (\epsilon^{A}, p_A) $ is dominated by the M1 multipole, we restrict ourselves to the single amplitude with $\alpha_1 = 1$, $\alpha_2 = \beta_1 = \beta_2=0$. 
The helicity amplitudes $\mathcal{M}_{\lambda_\gamma \lambda_A}$ in the tensor rest frame are: 
\bsub\begin{align}
    \mathcal{M}_{11} & = \frac{p}{\sqrt{6}},
    &
    \mathcal{M}_{10} & = \frac{E_A p}{\sqrt{2}m_A} ,
    \\
    \mathcal{M}_{1-1} & = p ,
    &
    \mathcal{M}_{-1, \lambda_A}&= - \mathcal{M}_{1,- \lambda_A},
\end{align}\esub
times the overall coupling $\beta^{\gamma T}_A$. 
We write the amplitude of the process $\gamma(\lambda_\gamma) \gamma(\lambda_\gamma') \to T(\lambda_T)T(\lambda_T')$ with axial exchange in the $t$ channel at leading order in $s$. By matching to the expected form 
[Eq.~\eqref{eq:7}]:
\begin{align}
    A_{\lambda_\gamma\lambda_T,\lambda_\gamma'\lambda_T'} & = -\left(\beta_{A}^{\gamma T}\right)^2 \beta^A_{\lambda_\gamma,\lambda_T} 
    \left(\frac{-t}{m_T^2}\right)^{\frac{1}{2}|\lambda_\gamma-\lambda_T|} \frac{s}{m_A^2-t} \nonumber\\
    &\quad\times\left(\frac{-t}{m_T^2}\right)^{\frac{1}{2}|\lambda_\gamma'-\lambda_T'|}
    \beta^A_{-\lambda_\gamma',-\lambda'_T},
\end{align}
we get the structures in Table~\ref{tab:TVV}.

\section{Isospin relations}
\label{app:isospin} 
The transition of tensor to axial mesons is dominated by the M1 multipole. In the quark model, this requires a spin flip from $S=1$ to $S=0$ to conserve charge conjugation. In the tensor rest frame, the matrix elements read~\cite{Halzen:1984mc,LeYaouanc:1988fx}:
\begin{align}
   \mathcal{M}(T\to \gamma A) &\propto \sum_{i=1,2} \left\langle A, \lambda_A \big| \mu_i  \,\vec \sigma_i \cdot \vec {\epsilon^*}(\lambda_\gamma) \big| T, \lambda_T\right\rangle ,
\end{align}
where the sum runs over the two quarks, $\mu_i$ is the quark magnetic moment, $\sigma_i$ the spin operator, and $\epsilon$ the emitted photon polarization.

The transition of tensor to vector meson is instead dominated by the E1 multipole, and does not involve the quark spin:
\begin{align}
  &\mathcal{M}(T\to \gamma V) \propto \int \frac{d^3 p_1}{(2\pi)^3}\frac{d^3 p_2}{(2\pi)^3} (2\pi)^3 \delta^3\!\left(\vec p_1 + \vec p_2\right) \nonumber\\
   &\qquad\times\sum_{i=1,2}\left\langle V, \lambda_V \big| e_i \vec {\epsilon^*}(\lambda_\gamma) \cdot \vec p_i  \big| T, \lambda_T\right\rangle,
\end{align}
where $\vec p$ is the 3-momentum of the quark in the center-of-mass frame.

We align the spin quantization axis along the direction of the emitted photon. We consider a right-handed photon, and the tensor helicity $+2$. The wave functions are: 
\bsub\begin{align}
    |T \rangle & = \frac{1}{\sqrt{2}} (|u \bar u\rangle \mp |d \bar d\rangle)\,\, |\!\uparrow \uparrow\rangle\, R_{1,1}(r) Y^1_1(\theta,\phi),\\
    |A \rangle & = \frac{1}{2} (|u \bar u\rangle \mp |d \bar d\rangle)\,(|\!\uparrow \uparrow\rangle - |\!\downarrow \downarrow\rangle)\, R_{1,1}(r) Y^1_1(\theta,\phi),\\
    |V \rangle & = \frac{1}{2} (|u \bar u\rangle \mp |d \bar d\rangle)\,(|\!\uparrow \uparrow\rangle + |\!\downarrow \downarrow\rangle)\, R_{1,0}(r) Y^0_0(\theta,\phi),
\end{align}\esub
where $Y^m_\ell(\theta,\phi)$ are the usual spherical harmonics, and $R_{n,\ell}(r)$ the (unspecified) radial functions.
The upper (lower) sign is for isovector (isoscalar) mesons.
We are implicitly assuming that the orbital wave function of tensors and axials is the same. So are the isovector and isoscalar wave functions. We are also assuming ideal mixing for the mesons, namely that no strange component is included in the wave functions.
\begin{figure*}
	\includegraphics[width=0.32\textwidth]{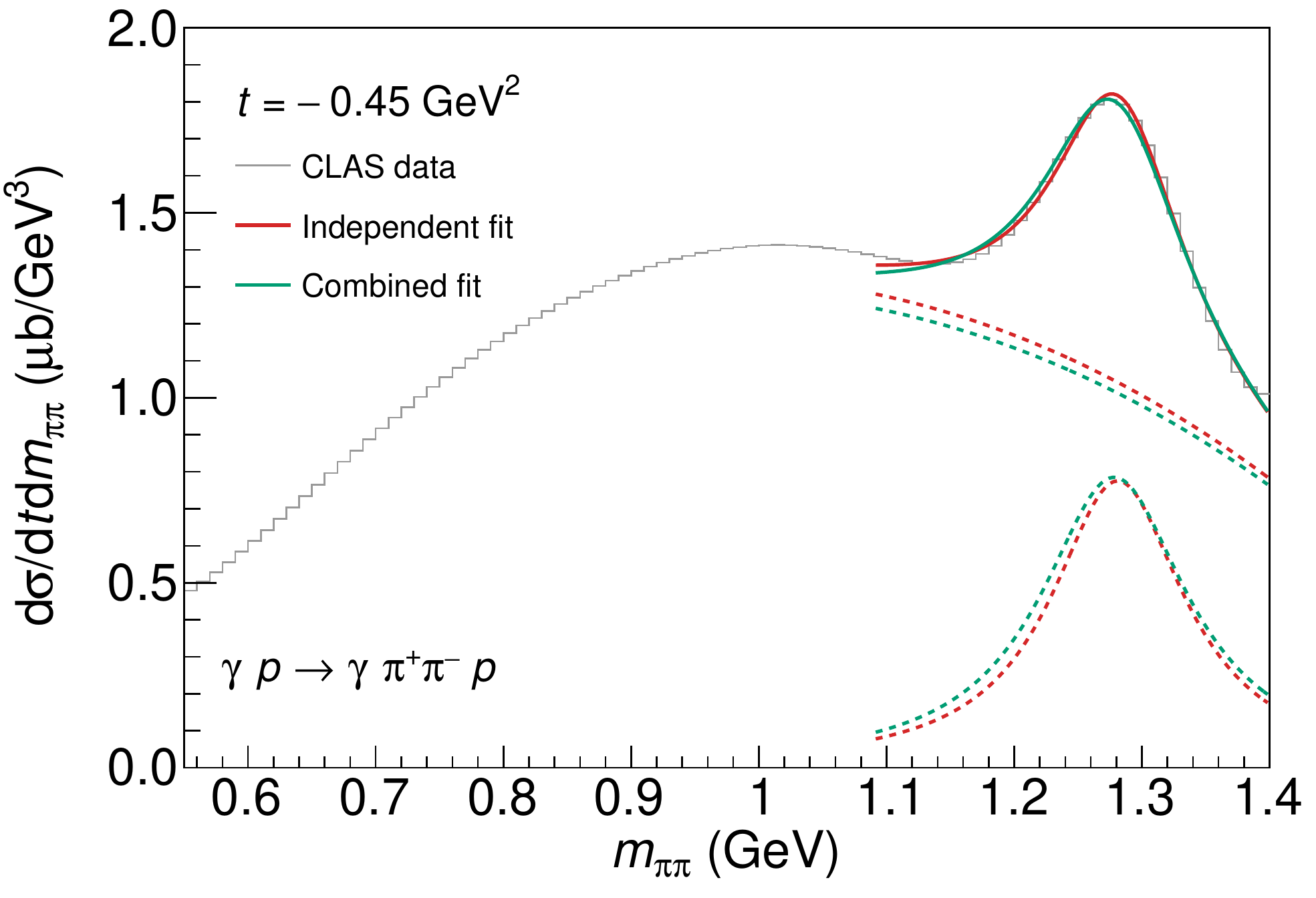}
	\includegraphics[width=0.32\textwidth]{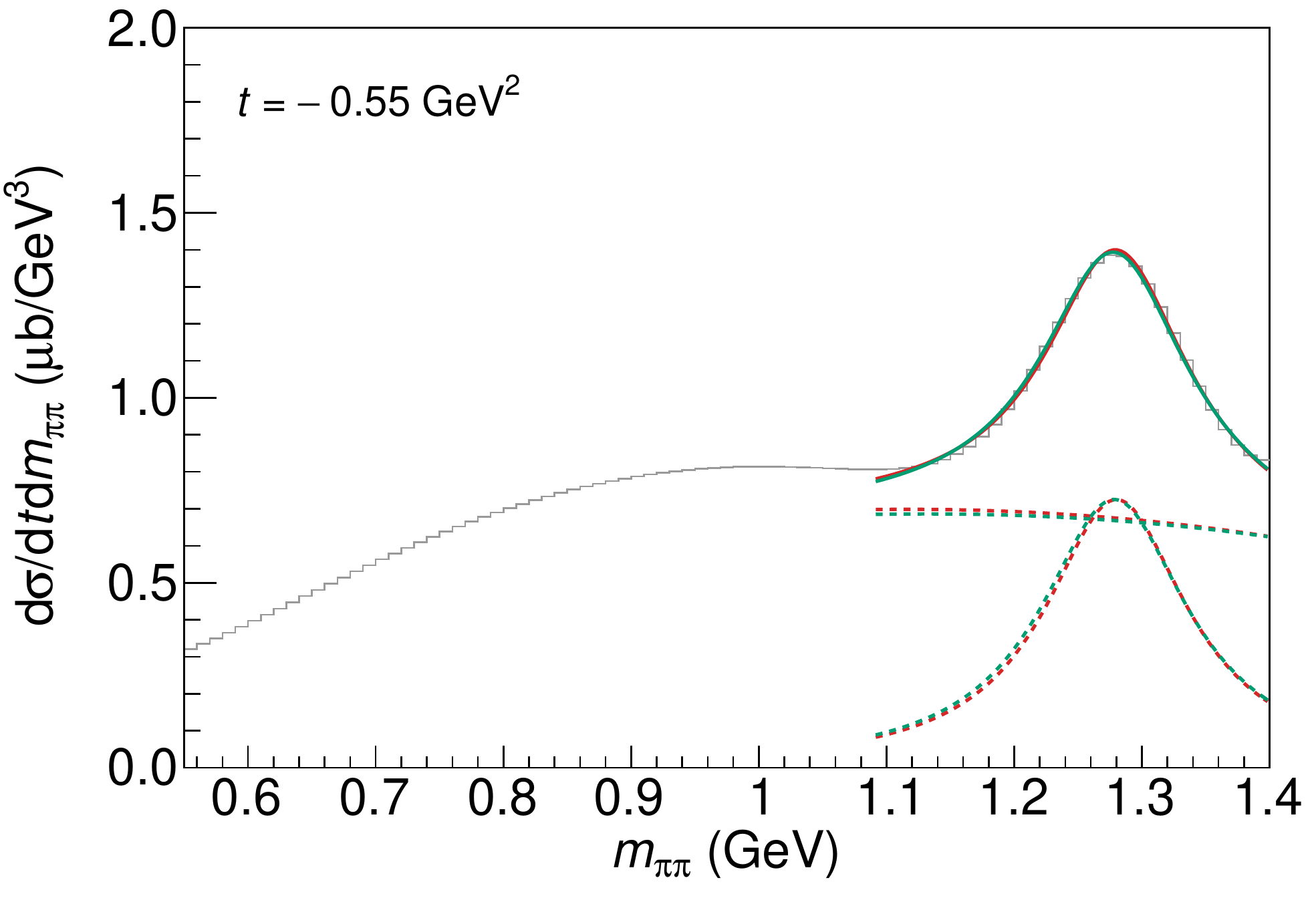}
	\includegraphics[width=0.32\textwidth]{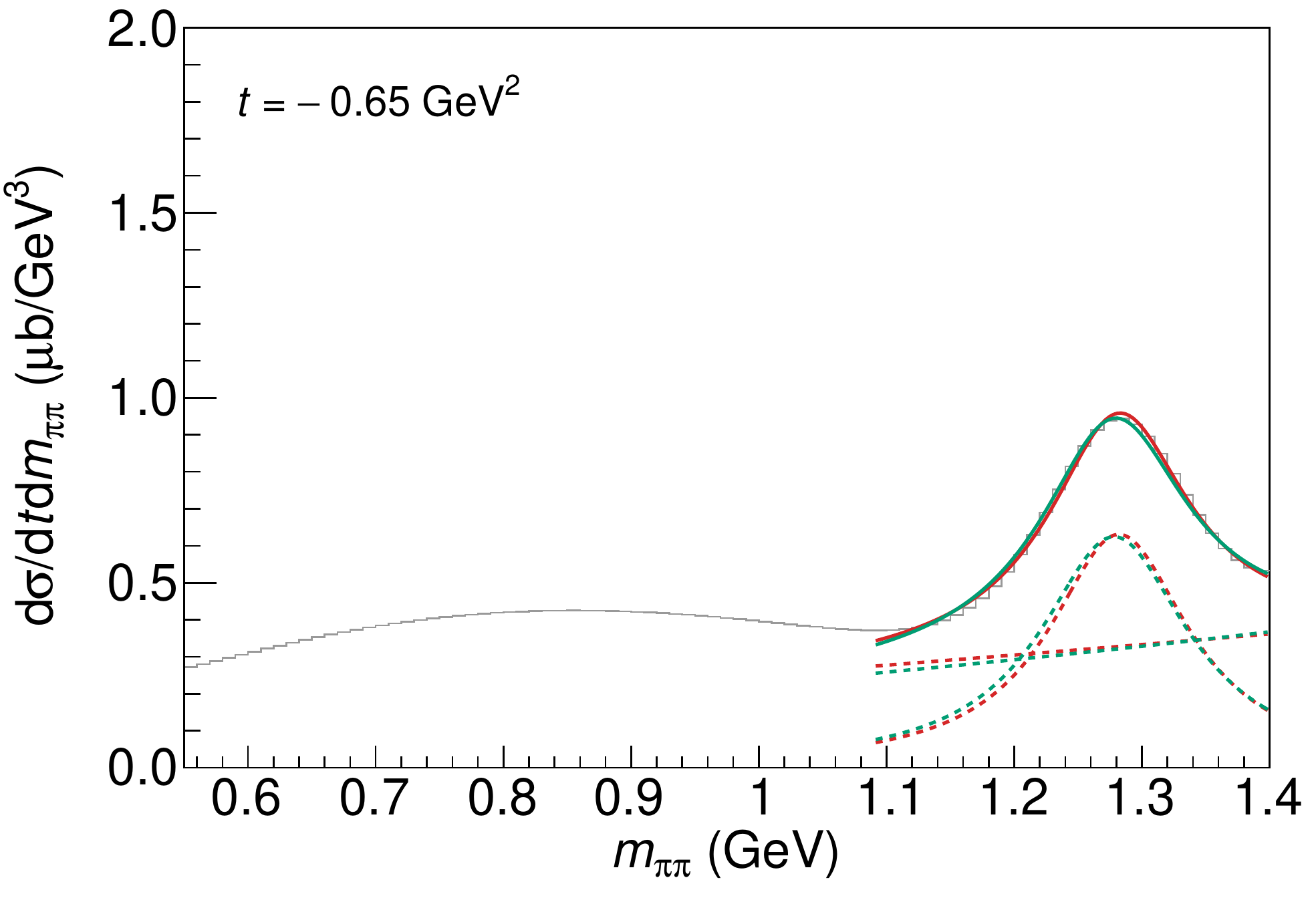}\\
	\includegraphics[width=0.32\textwidth]{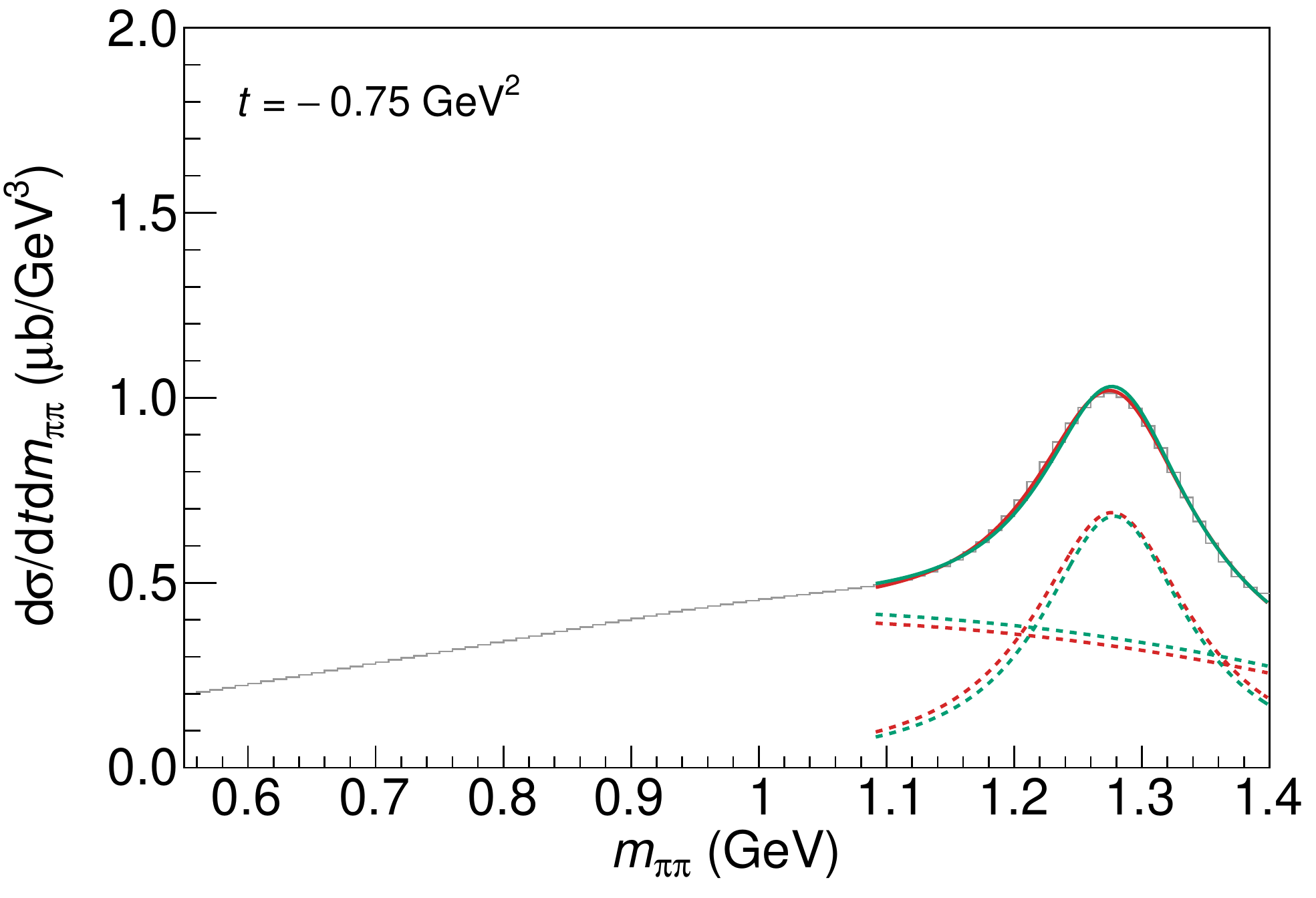}
	\includegraphics[width=0.32\textwidth]{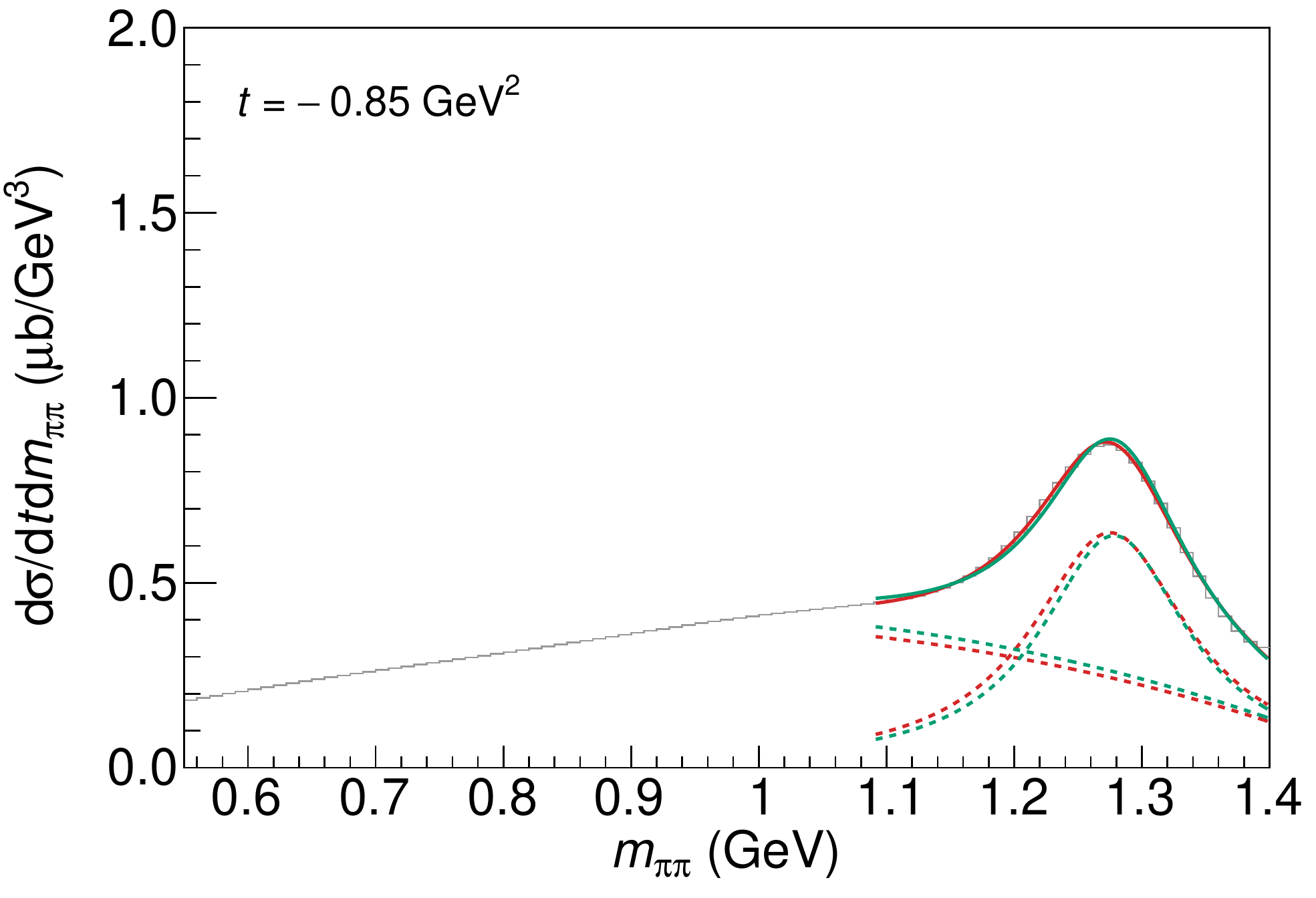}
	\includegraphics[width=0.32\textwidth]{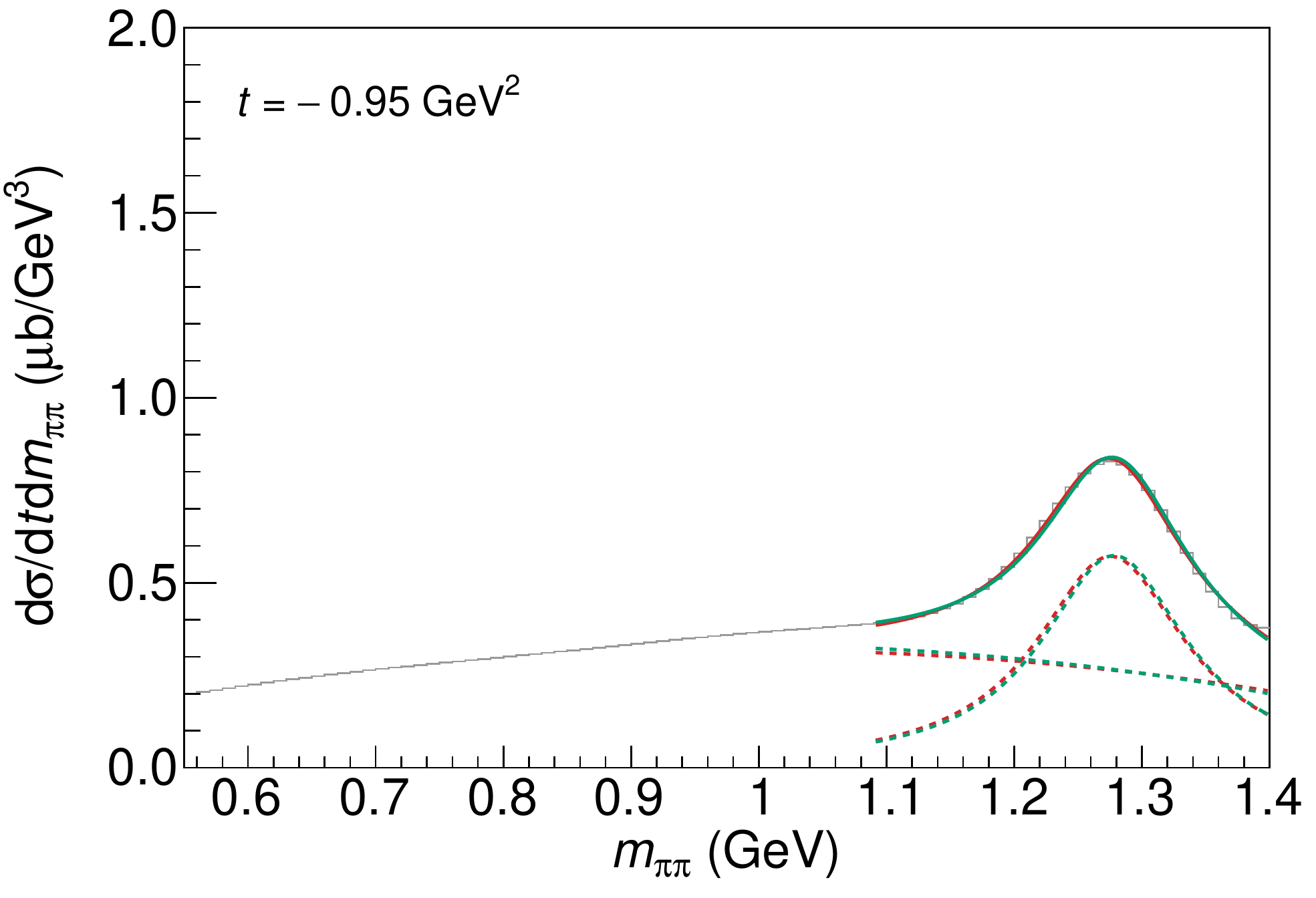}
	\caption{\label{fig:f2fit}Fit to the CLAS $D$-wave data on $\pi^+\pi^-$ photoproduction, as discussed in the text. Data are averaged over the four beam energy bins, $E_\gamma = 3.0$--$3.8\gev$. Mass and width of $f_2$ are fitted independently in each $t$ bin (red curve), or constrained to be the same (green curve). In dashed lines we show the separate contributions of $f_2$ and of the linear background. The strength of the $f_2$ looks constant in $t$, while the strength and shape of background changes dramatically.}
\end{figure*}

The magnetic moment is proportional to the electric charge $\mu_i = e_i/2m$ where $m$ is the constituent light quark mass:
\begin{subequations}
\begin{align}
  \mathcal{M}(f_2\to \gamma b_1) &= \mathcal{M}(a_2\to \gamma h_1) \propto (e_u - e_d)\frac{1}{2m}\\
  \mathcal{M}(f_2\to \gamma h_1) &= \mathcal{M}(a_2\to \gamma b_1) \propto (e_u + e_d)\frac{1}{2m}\\
  \mathcal{M}(f_2\to \gamma \rho) &= \mathcal{M}(a_2\to \gamma \omega) \propto \left(e_u - e_d\right) I\\
  \mathcal{M}(f_2\to \gamma \omega) &= \mathcal{M}(a_2\to \gamma \rho) \propto \left(e_u + e_d\right) I,
\end{align}
\end{subequations}
with $I = \int r^2 dr  R_{1,0}(r) \left(-i \partial_r \right) R_{1,1}(r)$. Since $e_u - e_d = \sqrt{4\pi \alpha}$, and $e_u + e_d = \sqrt{4\pi \alpha}/3$, we get:
\bsub\begin{align}
    \betaomegaa & = 
    \betarhof = 
    3\betarhoa =
    3\betaomegaf,
    \\
    \betabf & =
    \betaha
    =
    3\betaba =
    3\betahf.
\end{align}\esub

The two photons couplings become:
\bsub\begin{align}
   \beta_{f_2}^{\gamma\gamma} & =  \sqrt{4\pi\alpha}\left(\betarhof \frac{f_\rho}{m_\rho} + \betaomegaf \frac{f_\omega}{m_\omega} \right) \nonumber\\
       &=  \sqrt{4\pi\alpha}\,\betarhof \left( \frac{f_\rho}{m_\rho} +  \frac{1}{3}\frac{f_\omega}{m_\omega} \right),\\
   \beta_{a_2}^{\gamma \gamma} &= \sqrt{4\pi\alpha}\left(\betarhoa \frac{f_\rho}{m_\rho} + \betaomegaa \frac{f_\omega}{m_\omega} \right)\nonumber\\
          &=  \sqrt{4\pi\alpha}\,\betaomegaa \left(\frac{1}{3} \frac{f_\rho}{m_\rho} +  \frac{f_\omega}{m_\omega} \right),
\end{align}\esub
that are used in Eqs.~\eqref{eq:fromgammatoomega} and~\eqref{eq:fromgammatorho}.
For the decay constants, following the arguments above one gets $f_\rho = 3f_\omega$. This relation is broken at the $10\%$ level, suggesting some contributions from annihilation diagrams neglected here. If we apply this relation, and set $m_\rho = m_\omega$, we get 
$\beta_{a_2}^{\gamma\gamma} = \frac{3}{5} \beta_{f_2}^{\gamma\gamma}$, in agreement with the experimental values.

\section{Extraction of $f_2(1270)$ cross section}
\label{app:f2xsect}



\begin{figure}[b]
	\includegraphics[width=\columnwidth]{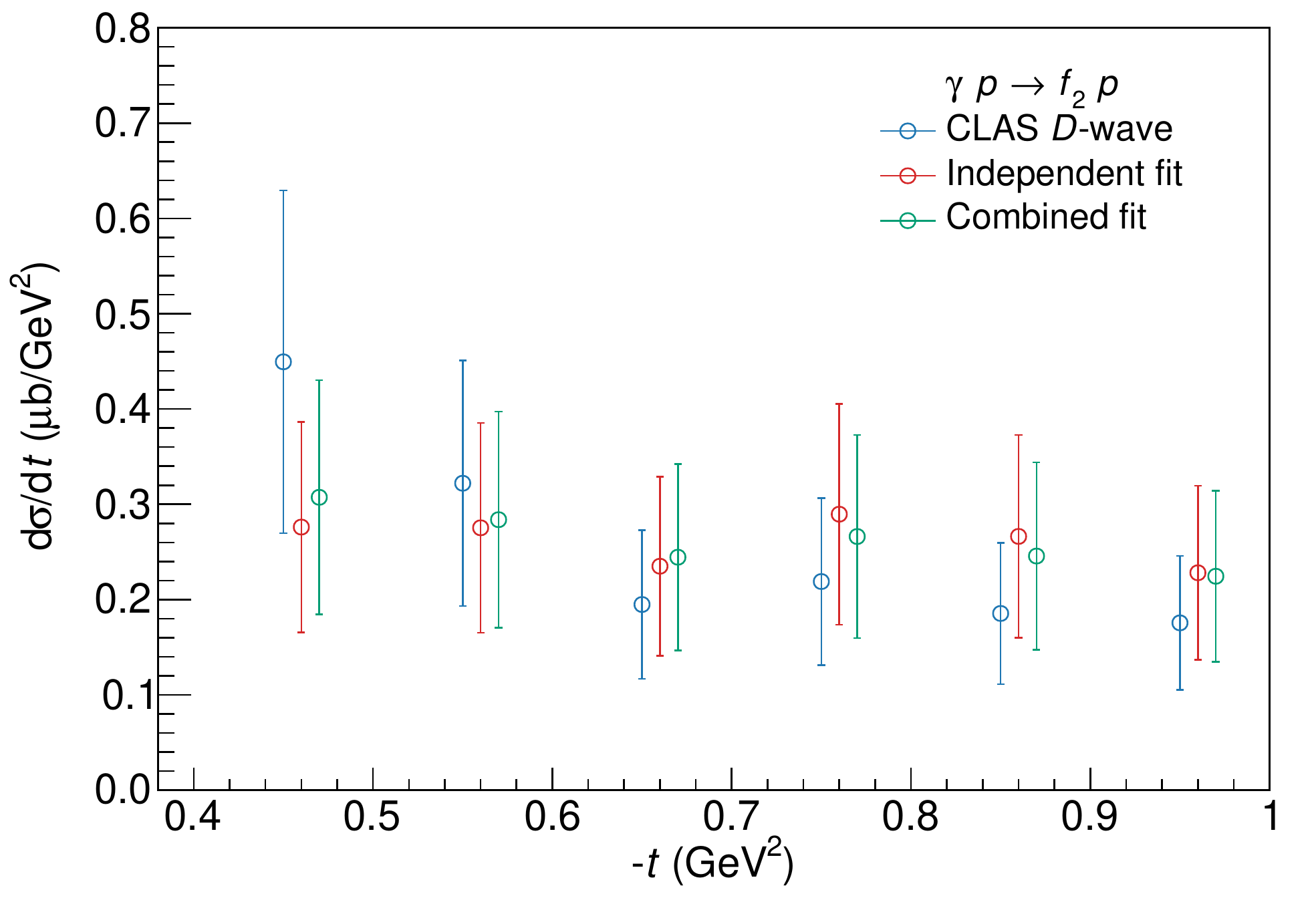}
	\caption{\label{fig:f2data} Differential cross section of $f_2$. A $40\%$ systematic error is shown. We compare with the CLAS data points from Fig.~24 of~\cite{Battaglieri:2009aa} (blue). We remind that CLAS points were obtained by integrating the $m_{\pi\pi}$ bins in the $[1090,1400]\mev$ range, that roughly corresponds to $[m_{f_2} - \Gamma_{f_2},m_{f_2} + \frac{2}{3}\Gamma_{f_2}]$. Some $f_2$ signal is lost, and a substantial background is included. Moreover, the branching ratio $\mathcal{B}\!\left(f_2\to \pi\pi\right)$ is not included. The red and green points correspond to the different extractions, namely if the $f_2$ mass and width is fitted independently or not in the different $t$ bins. These two results are consistent within error. Red and green points are slightly shifted horizontally to ease the reading.}
\end{figure}

As we mentioned, CLAS published the partial wave analysis of $\pi^+\pi^-$ photoproduction for $3.0$--$3.8\gev$ beam energy range~\cite{Battaglieri:2009aa}. The $t$ dependence of the $f_2$ was not directly extracted. The plot in Fig.~24 of~\cite{Battaglieri:2009aa} indeed reports the differential cross section integrating the $\pi\pi$ invariant mass over the $f_2$ peak region, $m_{\pi\pi}\in [1090, 1400]\mev$. This would be a good estimate for the $f_2$ differential cross section if the background underneath the peak were negligible. One can appreciate from Fig.~14  of~\cite{Battaglieri:2009aa}  that this is not the case. The published version of the paper does not report the $\pi\pi$ invariant mass in bins of $t$. However, the $D$-wave dataset can be downloaded from the HEPDATA repository, in bins of $t$ and beam energy~\cite{hepdatabatta}. We see that the amount of background is even larger at small values of $t$. We extract the $f_2$ cross section by fitting the $D$-wave data in the $f_2$ region with a simple constant width Breit-Wigner on top of a incoherent linear background:
\begin{align}
    &\frac{\diff \sigma\left(\gamma p \to \left(\pi^+\pi^-\right)_\text{$D$-wave} p\right)}{\diff t \: \diff m_{\pi\pi}} = 2m_{\pi\pi} \nonumber\\
    &\quad\times \Bigg[     \frac{\diff \sigma\left(\gamma p \to f_2 p\right)}{\diff t}\frac{1}{\pi} \frac{m_{f_2} \Gamma_{f_2} \mathcal{B}\!\left(f_2\to \pi^+\pi^-\right)}{\left(m_{f_2}^2 - m_{\pi\pi}^2\right)^2 + m_{f_2}^2 \Gamma^2_{f_2}}\nonumber\\
    &\qquad + c m_{\pi\pi}^2 + d \Bigg],
\end{align}
where $\mathcal{B}\!\left(f_2\to \pi^+\pi^-\right) = 56.2^{+1.9}_{-0.6}\%$~\cite{pdg}. We fit to the $1.09$-$1.4$ $m_{\pi\pi}$ range only, in order to have an easier description of the background. Since the error quoted in HEPDATA are systematic only, we ignore them in the fit, assuming equal weights for each bin, and assign a $40\%$ error to our final results, consistently with what done in~\cite{Battaglieri:2009aa}. Data are available in for also bins of beam energy, from $3.0$ to $3.8\gev$, but the energy dependence of data is mild. Therefore, we average data over the four bins, and quote the results at the mean energy $E_\gamma = 3.4\gev$. The fit is shown in Fig.~\ref{fig:f2data}. The $f_2$ mass and width is fitted independently in the six $t$ bins, obtaining results consistent with each other, and with the PDG value. Alternatively, we impose mass and width to be the the same in all $t$ bins. The final result is the same within errors, as seen in Fig.~\ref{fig:f2data}. We notice that the backgrounds depends on $t$ much more than the $f_2$.

\section{$f_2(1270)$ couplings} \label{app:f2}
We determine the \betaVf couplings from the decay width:
\begin{align}
    &\Gamma(f_2\to \rho^0\rho^0 + \rho^+\rho^-) \nonumber \\
    &\quad\simeq \Gamma(f_2\to 2\pi^+ 2\pi^- + \pi^+\pi^- 2\pi^0)  =19.6^{+4.0}_{-8.6}\mev,
\end{align}
assuming that the pion system is saturated by $\rho$ mesons.

The matrix element $\sum_\text{pol}|\mathcal{M}|^2$ given by Eqs.~\eqref{width:TVV:Min} and \eqref{width:TVV:TMD} must be averaged over the two $\rho$ line shapes:
\begin{align} \nonumber
    &\Gamma(f_2\to \rho^0\rho^0 + \rho^+\rho^-) = \frac{3}{2}\frac{\left(\beta_{f_2}^{\rho\rho}\right)^2}{40\pi m_{f_2}^4}  \int\!\int \frac{\diff s'}{\pi} \frac{\diff s''}{\pi} \sum_\text{pol}|\mathcal{M}|^2\\ &\times \frac{\lambda^{1/2}\left(m_{f_2}^2, s', s''\right)}{2m_{f_2}}
B_\rho(s') B_\rho(s'') \,
\theta(\lambda\!\left(m_{f_2}^2, s', s''\right)),
\end{align}
where $B_\rho(s)$ is given in Eq.~\eqref{eq:rhols}, and the factor of $3/2$ takes into account the sum over isospin and the identical particle phase space.
VMD allows us to get:
\begin{align}
    \betarhof & = \sqrt{4\pi \alpha} \frac{f_\rho}{m_\rho} \beta_{f_2}^{\rho\rho}.
\end{align}

Alternatively, the \betarhof can be extracted from the two-photon width,     $\Gamma(f_2\to \gamma \gamma)  = 2.6 \pm 0.5\kev$~\cite{pdg} and  Eqs.~\eqref{eq:2photon_min} and \eqref{eq:2photon_TMD}, to extract the two-photon couplings $\beta_{f_2}^{\gamma\gamma}$ for the two models. We then obtain the \betaVf couplings from:
\begin{align}
    \betarhof & = \frac{\beta_{f_2}^{\gamma\gamma}}{\sqrt{4\pi \alpha}} \left( \frac{f_\rho}{m_\rho} + \frac{1}{3} \frac{f_\omega}{{m_\omega}} \right)^{-1},
    \label{eq:fromgammatorho}
\end{align}
derived in Appendix~\ref{app:isospin} within the quark model. The coupling to $\omega$ can be obtained from either determination, using:
\begin{equation}
        \betaomegaf= \frac{1}{3}\betarhof.
\end{equation}
The numerical values under the different assumptions are summarized in Table~\ref{tab:param}.

\section{Spin and polarization observables}
\label{app:obs}
Experimentally, observables related to tensor meson photoproduction are extracted from their 
decay products. The simplest final state to detect is  
two pseudoscalars, \ie $\eta\pi$ for $a_2$ and $\pi\pi$ for $f_2$. The general case of two pseudoscalar photoproduction with a linearly polarized beam has been treated in detail in~\cite{Mathieu:2019fts}. We summarize here the relevant formulae when the tensor meson is so narrow that the existence of other partial 
waves can be neglected.

For a linearly polarized photon, the differential cross section is:
\begin{align}\nonumber
&I(\Omega,\Phi)  = \frac{\diff \sigma\!\left(\gamma p \to T(\to PP') \,p \right)}{\diff t \: \diff \Omega \: \diff \Phi }
\\&\quad =\kappa
\sum_{ \substack{\lambda_\gamma\lambda'_\gamma \\ \lambda_p\lambda'_p}} \mathcal{A}_{\lambda_\gamma; \lambda_p\lambda'_p} (\Omega) \rho^\gamma_{\lambda_\gamma\lambda'_\gamma}(\Phi) \A{\lambda'_\gamma}{\lambda_p\lambda'_p}^* (\Omega),
\label{eq:xsecOmega}
\end{align}
where $\Phi$ is the azimuthal angle between the 
polarization plane (which contains the photon polarization and momentum) and the production plane (which contains the photon, tensor and recoiling proton momenta), while $\Omega =(\theta , \phi)$ are the decay angles of the pseudoscalar $P$ in the helicity frame. 
The photon spin density matrix elements (SDME) are  $\rho^\gamma_{\lambda_\gamma \lambda'_\gamma}(\Phi) = \tfrac{1}{2}\left[\mathds{1}- P_\gamma \left(\sigma^1 \cos 2\Phi + \sigma^2 \sin 2 \Phi\right)\right]_{\lambda_\gamma \lambda'_\gamma}$, with $P_\gamma$ the beam polarization, and $\sigma^{1,2}$ the Pauli matrices. In the narrow width approximation, $\A{\lambda_\gamma}{ \lambda_p\lambda'_p}\A{\lambda'_\gamma}{\lambda_p\lambda'_p}^* \propto \delta\!\left(m^2_{P P'} - m_T^2\right)$, and the dependence of $\mathcal{A}$ on $s,t$ is understood. We include all numerical factors in:
\begin{align} \label{eq:pspace}
\kappa & = \frac{1}{2}\frac{1}{16\pi} \frac{1}{2\pi} \frac{1}{(2m_p E_\gamma)^2}\times \left\{
\begin{matrix}\frac{1}{2} & \text{for $f_2 \to \pi^0\pi^0$}\\1 & \text{otherwise}\end{matrix}\right..
\end{align}
The amplitude is saturated by the $D$ wave:
\begin{align}
 \A{\lambda_\gamma}{\lambda_p\lambda'_p} = \sum_m \M{\lambda_\gamma m}{\lambda_p \lambda'_p} Y^m_2(\Omega).
\end{align}

With a linearly polarized beam, only two observables are accessible when the decay angles are integrated over; 
the differential cross section $\diff \sigma/\diff t$ and the integrated beam asymmetry $\Sigma_{4 \pi}$:
\begin{align} \label{eq:Sigma4pi}
\frac{\diff \sigma}{\diff t \: \diff \Phi} & = \frac{1}{2\pi}\frac{\diff\sigma}{ \diff t}\left(1 + P_\gamma \Sigma_{4\pi} \cos 2 \Phi \right),
\end{align}
where:
\bsub\begin{align}
 \frac{\diff\sigma}{\diff t}  & = \pi \kappa \sum_{ \substack{\lambda_\gamma m \\ \lambda_p\lambda'_p}}
 \left| \M{\lambda_\gamma m}{\lambda_p \lambda'_p}\right|^2 \equiv 2\pi \kappa N,
 \\
 \Sigma_{4\pi} & = -\frac{1}{2N}\sum_{\substack{\lambda_\gamma m \\ \lambda_p\lambda'_p}}
  \M{-\lambda_\gamma m}{\lambda_p \lambda'_p} \mathcal{M}^*_{\lambda_\gamma m ; \lambda_p \lambda'_p}.
\end{align}\esub
The component proportional to $\sin 2\Phi$ vanishes indeed upon integration over $\Omega$ because of parity conservation.
The angular dependence allows one 
to extract the SDME, defined as:
\bsub\begin{align}
 \rho^0_{mm'} & = \frac{1}{2N} \sum_{\lambda_\gamma,\lambda^{(\prime)}_p} \M{\lambda_\gamma m}{\lambda_p \lambda'_p} \mathcal{M}^*_{\lambda_\gamma m'; \lambda_p \lambda'_p}, 
 \\
  \rho^1_{mm'} & = \frac{1}{2N} \sum_{\lambda_\gamma,\lambda^{(\prime)}_p} \M{-\lambda_\gamma m}{\lambda_p \lambda'_p} \mathcal{M}^*_{\lambda_\gamma m'; \lambda_p \lambda'_p},
  \\
   \rho^2_{mm'} & = \frac{i}{2N} \sum_{\lambda_\gamma,\lambda^{(\prime)}_p} \lambda_\gamma \M{-\lambda_\gamma m}{\lambda_p \lambda'_p} \mathcal{M}^*_{\lambda_\gamma m'; \lambda_p \lambda'_p}.
\end{align}\esub
They satisfy $[\rho^\alpha_{mm'}]^* = \rho^\alpha_{m'm}$. Parity conservation implies:
\bsub\begin{align}
	\rho^0_{-m-m'}  & = \phantom{-}(-1)^{m-m'}\rho^0_{mm'},
	\\
	\rho^1_{-m-m'}  & = \phantom{-}(-1)^{m-m'}\rho^1_{mm'},
	\\
	\rho^2_{-m-m'}  & = -(-1)^{m-m'}\rho^2_{mm'}.
\end{align}\esub
The SDME are normalized such that:
\bsub\begin{align}
\rho^0_{00} + 2\rho^0_{11} +2 \rho^0_{22} &= 1,\\
\rho^1_{00} + 2\rho^1_{11} + 2\rho^1_{22} &= - \Sigma_{4\pi}.
\end{align}\esub

We use the reflectivity basis~\cite{Mathieu:2019fts}. The SDME can be split into reflectivity components using
\begin{align}
 \rho^{(\pm)}_{mm'} & = \frac{1}{2} \left( \rho^0_{mm'} \mp (-1)^{m'} \rho^1_{m-m'} \right).
\end{align}
The convention is such that the natural (unnatural) exchanges contribute only to $\rho^{(+)}_{mm'}$ ($\rho^{(-)}_{mm'}$) at the leading order in the energy squared~\cite{Mathieu:2019fts}. 

We  decompose the intensity~\eqref{eq:xsecOmega} as:
\begin{align}  \nonumber
I(\Omega,\Phi)
	& =\frac{5}{4\pi} \frac{1}{2\pi} \frac{\diff \sigma}{\diff t}  \big[ W^0(\Omega) \\
	& \quad - W^1(\Omega)  P_\gamma \cos \Phi  - W^2(\Omega)  P_\gamma \sin \Phi \big].
	\label{eq:intensityW}
\end{align}
The SDME can be extracted from the angular dependence of the intensities:
\begin{widetext}
\begin{align} \nonumber
 W^\alpha(\Omega) & =  \frac{1}{16} \rho^\alpha_{00} (1+3\cos2\theta)^2  
 - \frac{3}{4} \rho^\alpha_{1-1} \sin^2 2 \theta \cos 2 \phi 
 - \sqrt{\frac{3}{8}} \re \rho^\alpha_{10}\sin 2 \theta (1+3 \cos2 \theta) \cos \phi 
 \\ \nonumber
 &    +\frac{3}{4} \rho^\alpha_{11}  \sin^2 2\theta
 + 3 \re \rho^\alpha_{2-1}\cos \theta \sin^3\theta \cos3 \phi
 + \frac{3}{4}\rho^\alpha_{2-2} \sin^4 \theta \cos 4 \phi
 \\ \nonumber
 &  + \sqrt{\frac{3}{8}} \re \rho^\alpha_{20} (1+3 \cos2 \theta) \sin^2\theta \cos 2 \phi 
 - 3 \re \rho^\alpha_{21}\cos \theta \sin^3\theta \cos \phi  + \frac{3}{4} \rho^\alpha_{22} \sin^4 \theta,
\end{align}
valid for $\alpha = 0,1$. The intensity $W^2$ decomposes into:
\begin{align} \nonumber
W^2(\Omega) & = \sqrt{\frac{3}{8}} \im \rho^2_{10} \sin 2 \theta (1+3 \cos2\theta) \sin \phi+ 
 \frac{3}{4i} \rho^2_{1-1} \sin^2 2 \theta \sin 2\phi \\ \nonumber
 & - \sqrt{\frac{3}{8}} \im \rho^2_{20} \sin^2 \theta (1+3 \cos2\theta) \sin2 \phi 
 + 3\cos\theta \sin^3\theta  \left[  \im \rho^2_{21}  \sin\phi - \im \rho^2_{2-1}  \sin3\phi  \right] \\
 & - \frac{3}{4i} \rho^2_{2-2} \sin^4\theta \sin4 \phi.
\end{align}
\end{widetext}
We remind 
the reader that $\rho^{0,1}_{m\pm m}$ is purely real, $\rho^2_{m-m}$ purely imaginary, and $\rho^2_{00}=0$.

With a linearly polarized beam, the accessible reflectivity components are
$\rho^{(\pm)}_{m\pm m}$, and $\re\rho^{(\pm)}_{mm'}$.

Opposite reflectivities do not interfere, $\diff \sigma/\diff t = \diff \sigma^{(+)}/\diff t + \diff \sigma^{(-)}/\diff t$, with $\diff \sigma^{(\pm)}/\diff t = 2\pi \kappa N\left(\rho^{(\pm)}_{00} + 2\rho^{(\pm)}_{11} + 2\rho^{(\pm)}_{22}\right)$. 

The parity asymmetry:
\begin{align} \label{eq:Psig}
    P_\sigma & = \frac{ \frac{\diff \sigma^{(+)}}{\diff t} - \frac{\diff \sigma^{(-)}}{\diff t} }{\frac{\diff \sigma^{(+)}}{\diff t} + \frac{\diff \sigma^{(-)}}{\diff t}} & = 2 \rho^1_{1-1} -2 \rho^1_{2-2} -  \rho^1_{00},
\end{align}
measures the relative importance of the two reflectivity components. When the two pseudoscalar mesons only couple in a $D$-wave, $P_\sigma$ corresponds to the beam asymmetry along the $y$ axis, $\Sigma_y$, as defined in Ref.~\cite{Mathieu:2019fts}.


\bibliographystyle{apsrev4-2}
\bibliography{quattro}

\end{document}